\documentclass{aa}
\usepackage{natbib}
\usepackage{txfonts}
\usepackage{graphicx}

\newcommand{\gl}{$\lambda$}
\newcommand{\kms}{km\,s$^{-1}$}

\begin{document}

\title{Coronal versus photospheric abundances of stars \\ with different
  activity levels} 
\author{J. Sanz-Forcada$^1$ \and F. Favata$^2$ \and G. Micela$^1$}
\institute{INAF -- Osservatorio Astronomico di Palermo
G. S. Vaiana, Piazza del Parlamento, 1; Palermo, I-90134, Italy
  \and Astrophysics Division -- Research and Science Support Department
  of ESA, ESTEC, Postbus 299, NL-2200 AG Noordwijk, The Netherlands 
}
\offprints{J. Sanz-Forcada, \email{jsanz@astropa.unipa.it}}
\date{Received / Accepted}

\abstract{
  
We report a detailed analysis of the coronal abundance of 4 stars
with varying levels of activity and with accurately known
photospheric abundances. The coronal abundance is determined here
using a line flux analysis and a full determination of the
differential emission measure. Photospheric abundance values are
taken from literature works. Previous coronal abundance
determinations have generally been compared to solar photospheric
abundances; from this a number of general properties have been
inferred, such as the presence of a coronal metal depletion with an
inverse First Ionization Potential correlated with activity level.
Here we show that, when coronal abundances are
compared with real photospheric values for the individual stars, the
resulting pattern can be very different. Also, we present evidence
that, in some cases, the coronal metal abundance may not be uniform
in the corona; in particular it can vary with the
temperature of the emitting plasma.  }

   \maketitle
%
\section{Introduction}

Following the initial results from the ASCA-SIS instrument
\citep[e.g.][]{whi94,dra96}, which showed that the low-resolution
X-ray spectra of active coronal sources where often best fit with a
low metal abundance (typically about 0.3 the solar photospheric
abundance) the issue of the metal abundance of the coronal plasma has
become one of the main areas of interest for X-ray spectroscopy of
active stars.
Due to the limited spectral resolution of the ASCA spectra, the
determination of abundance for elements other than Fe was often not
accurate \citep{fav03}. The availability of the XMM-\emph{Newton} and
\emph{Chandra} high-resolution X-ray spectra of coronal sources has
allowed to observe individual, resolved spectral lines of a number of
elements in the corona of active stars, and thus to address the issue
of the abundance of a number of elements. Unfortunately up to now the
results have not been as conclusive as initially expected. One reason
is that often the same data analyzed with different approaches result
in significantly different value for the elements abundances
\citep[see][for a number of examples]{fav03}.

One of the issues most often addressed by coronal abundance studies is
whether a bias with the First Ionization Potential (FIP) of the
element is present. Such ``FIP effect'' is found in the solar
corona, where the abundance of elements with low FIP ($< 10$~eV; e.g.
Mg, Si, Fe) is observed to be enhanced, in the corona, with respect to
their photospheric values, while elements with high FIP ($\ga 10$~eV;
e.g.  O, Ne, Ar) remain at abundances similar to the photospheric
values. Such behavior is observed in the upper solar atmosphere at
temperatures higher than $\sim 1$~MK, both in the non-active regions
and in full-disk spectra \citep{lam95,feld00}.  EUVE observations
showed the same FIP bias to be present in stars with low activity
levels, such as $\alpha$~Cen~AB \citep[G2V+K1V,][]{drake97}, and
possibly (only limited knowledge of their photospheric abundance was
available at that time) $\xi$~Boo~A \citep[G8V,][]{lam99} and
$\epsilon$~Eri \citep[K2V,][]{lam96}.  EUVE spectra of more active
stars did not show any FIP effect, with however a rather low coronal
Fe abundance with respect to the solar photosphere, which suggested
some metal depletion in the corona of the active stars. \citet{feld00}
review some of the results on solar and stellar coronal abundances
previous to \emph{Chandra} and XMM-\emph{Newton}.

\begin{table*}
\caption{Observation log}\label{obslog}
\begin{center}
\begin{footnotesize}
\begin{tabular}{lccccrcrrrr}
\hline \hline
{Object} & {HD \#} & {SpT} & {Instrument} & {\gl \gl (\AA)} & {$t_{\rm exp}$ (ks)}
& {Date} & $d$ (pc) & $L_{\rm X}$ & 
$L_{\rm X}$/$L_{\rm bol}$ & $L'_{\rm X}$ \\ 
\hline
Procyon & 61421 & F5IV & \emph{Chandra}/LETG & 3--175 & 69.6 & 6 Nov 1999 & 3.50 &
8.89E27 & 3.39E-7 & 9.41E26\\
Procyon & 61421 & F5IV & \emph{Chandra}/LETG & 3--175 & 69.7 & 7 Nov 1999 & `` & ``
& `` & ``\\
$\epsilon$~Eri & 22049 & K2V & \emph{Chandra}/LETG & 3--175 & 105.3 & 21 Mar 2001 &
3.22 & 1.79E28 & 1.42E-5 & 9.65E27\\
$\lambda$~And & 222107 & G8III/? & \emph{Chandra}/HETG & 1.7--27 & 81.9 & 17
Dec 1999 & 25.8 & 2.74E30 & 3.25E-5 & 1.73E30\\
V851~Cen & 119285 & K2IV-III/? & XMM/RGS & 4--38 & 29.2 & 7 Mar 2003 & 76.2 &
6.46E30 & 3.01E-4 & 4.47E30\\
\hline
\end{tabular}
\end{footnotesize}
\end{center}
Note: Distances ($d$) from \citet{hippa}. 
$L_{\rm X}$ (erg s$^{-1}$)  in the range
5--100~\AA\ (0.12--2.4 keV), the range covered by the ROSAT/PSPC. 
$L'_{\rm X}$ (erg s$^{-1}$) in the
range 6--20~\AA\ (0.62--2.1 keV), common to RGS, HETGS and LETGS. 
\end{table*}

The new generation of X-ray satellites XMM-\emph{Newton} and
\emph{Chandra} offers access to high resolution spectra in a
wavelength region where numerous spectral lines from different
elements are formed at coronal temperatures. This permits to measure
the fluxes of individual lines, in principle allowing a much better
determination of the coronal thermal structure and abundances.
Initial results on coronal abundances determined from data from the
two observatories have been reviewed by \citet{dra02} and
\citet{fav03}. These new results seem to point towards a scenario with
stars with low activity levels showing, when their coronal and
photospheric abundances are compared, a solar-like FIP effect
\citep[case of $\alpha$~Cen,][]{raa03} or no FIP bias \citep[for
Procyon,][]{raa02}. Active stars (like HR~1099, II~Peg, AR~Lac or
AB~Dor), with a quite different coronal thermal structure, dominated
by material emitting at temperatures of $\sim 10$~MK, show a pattern
characterized by a general depletion of elements with low FIP (when
they are compared with solar photospheric values) while the elements
with high FIP remain at solar (or sometimes higher) photospheric
values \citep{dra01,bri01,huen01,huen03,sanz03} although photospheric
abundances of other elements than Fe are not known for HR~1099.
\citet{aud03} analyzed XMM-\emph{Newton} observations of 5 active
binary systems, concluding that low-FIP elements are subject to larger
depletion in more active stars, while high-FIP elements remain at
similar levels (although photospheric abundances were available only
for one of the objects in their sample, namely $\lambda$~And).  In
general, the comparison of stellar coronal abundances with solar
photospheric values is of little use, given the wide range of
abundances found among late type stars \citep[e.g.][]{cay01}.  Most of
the stars for which coronal abundances have been determined have no
published photospheric abundances, so that a true test of e.g.
FIP-related effects cannot be done. This is specially true for active
stars, for which the large rotational velocity complicates the
measurements of photospheric abundances. In most cases the available
photospheric abundances are limited to Fe \citep[e.g.][]{ran93}, and
even these in many cases are likely to be affected by systematic
errors.  Moreover, different methods to determine the distribution of
material with temperature in the corona (the so called ``emission
measure distribution'', EMD) sometimes produce contradictory results,
depending on the analysis technique employed \citep{fav03,sanz03}.

In present work we determine the coronal thermal structure and
abundances of four stars with different activity levels (in terms of
$L_{\rm X}/L_{\rm bol}$) observed with relatively high statistics with
\emph{Chandra} and XMM-\emph{Newton} (see Table~\ref{obslog}), and for
which there are accurate photospheric abundances measurements
available, with the aim to explore the relation between activity
levels and coronal relative abundance patterns. The chosen stars are
Procyon (one of the stars with lowest $L_{\rm X}/L_{\rm bol}$
values), the intermediate activity star $\epsilon$~Eri, the
long-period active binary $\lambda$~And and the active binary
V851~Cen. These four objects are among the few stars for which both
high-resolution coronal spectra with good statistics, and measurements
of photospheric abundances for several elements, are available.

The observations are described in Sect.~2, while a brief description
of the analysis techniques is presented in Sect.~\ref{sec:analysis}.
The scientific results for each of the star are discussed in
Sect.~\ref{sec:results}, followed by the discussion and conclusions in
Sect.~\ref{sec:conclusions}.

\begin{figure*}
\begin{center}
\includegraphics[angle=90,width=0.4\textwidth]{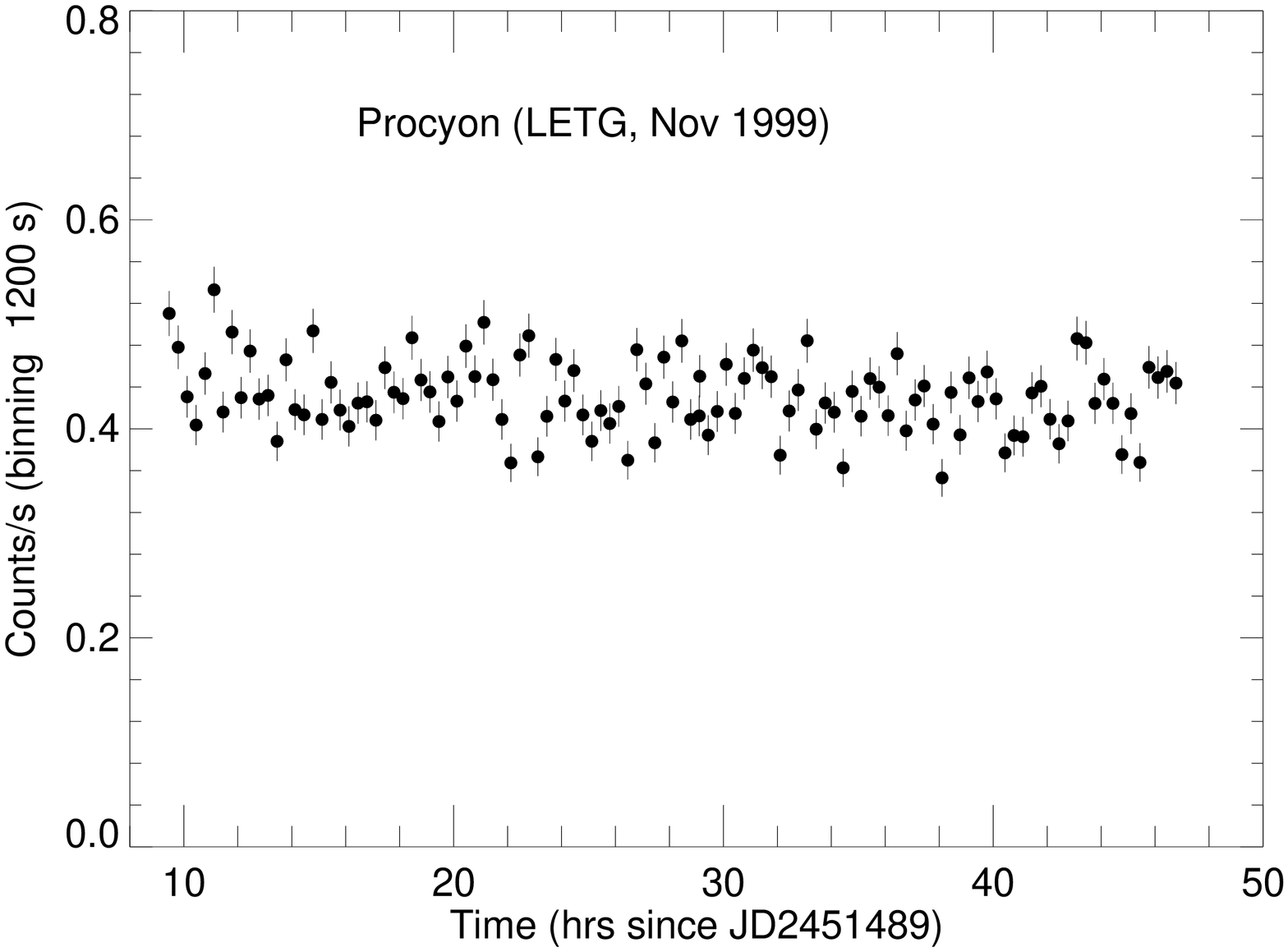}
\includegraphics[angle=90,width=0.4\textwidth]{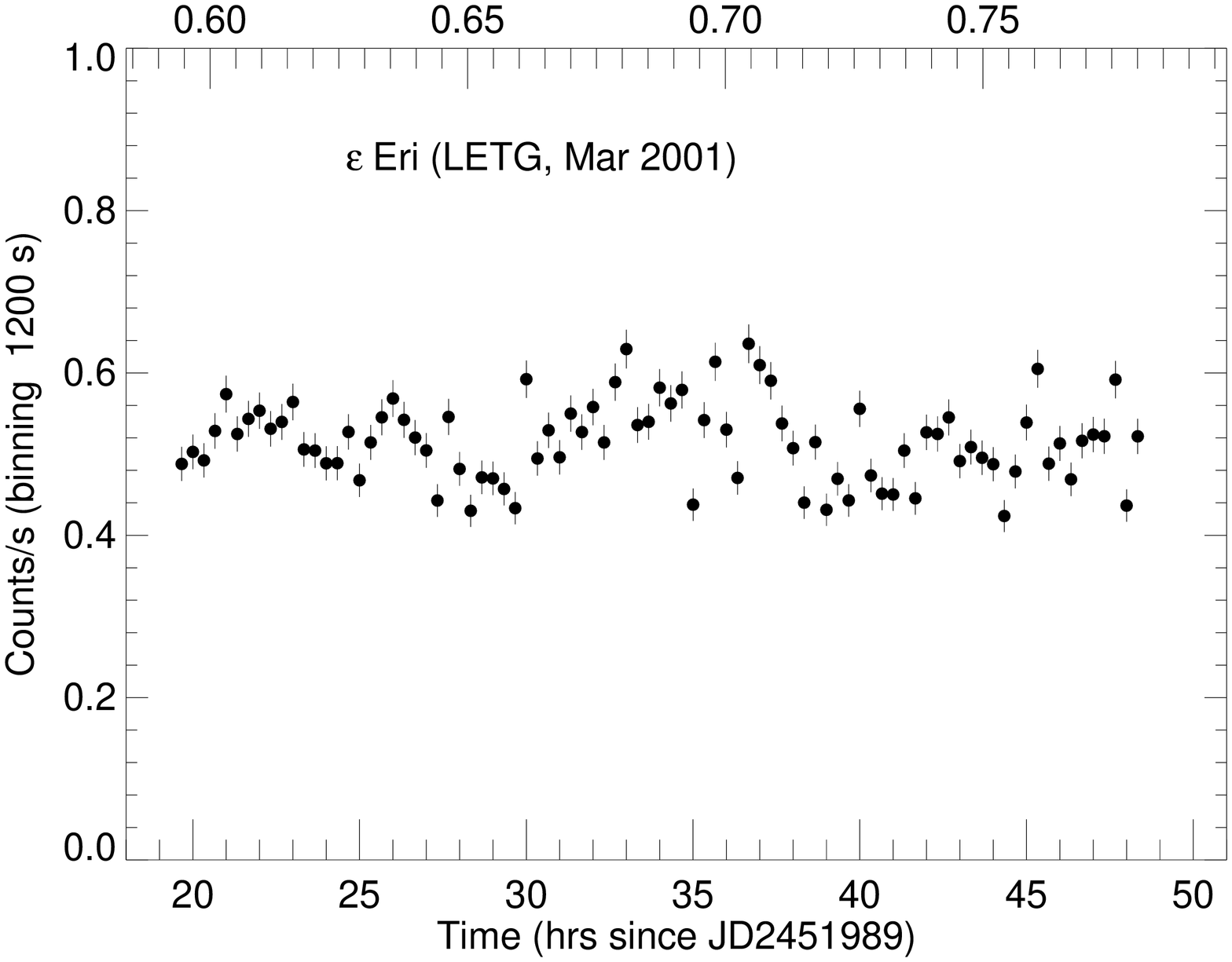}
\includegraphics[angle=90,width=0.4\textwidth]{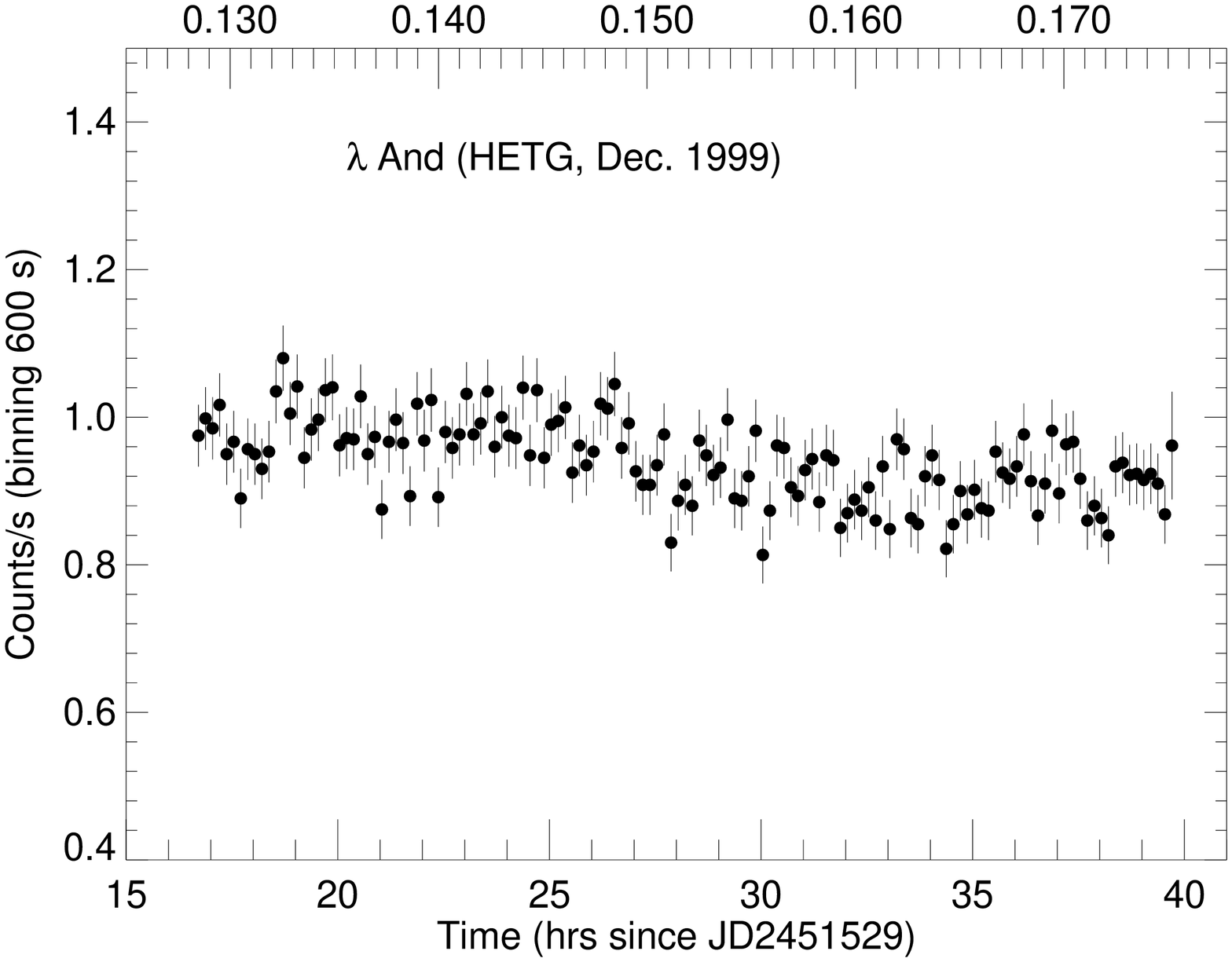}
\includegraphics[angle=90,width=0.4\textwidth]{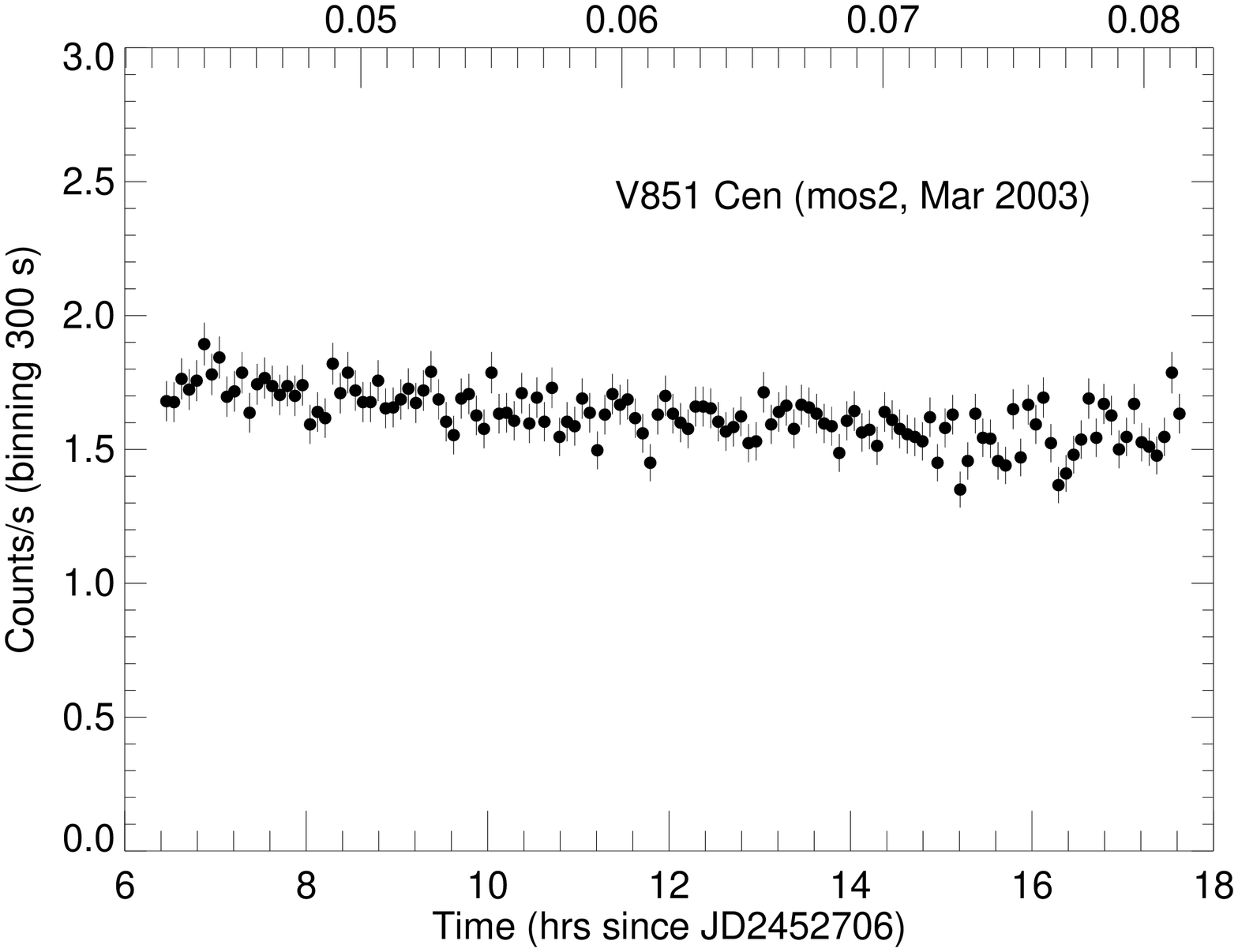}
\end{center}
 \caption{Light curves of Procyon (LETGS),  $\epsilon$~Eri (LETGS),
   $\lambda$~And (HETGS) and V851~Cen (XMM/MOS 2) with 1-$\sigma$
   error bars. The upper axis reports the photometric phase
   ($\epsilon$~Eri) and orbital phase ($\lambda$~And and V851~Cen,
   with $T_0$ corresponding to secondary star located behind the
   primary star).
   \label{lcurves}}
\end{figure*}

\section{Observations}

We have analyzed observations conducted using different X-ray
high-resolution spectrographs; however, the spectral lines observed in
the different spectra are formed in very similar temperature ranges
($\log T[{\rm K}]\sim 5.5$--$7.8$), hence yielding information of the
same region of the corona (see Sect.~\ref{sec:analysis}). The X-ray
luminosity of the targets has been calculated by modeling the observed
spectra, as listed in Table~\ref{obslog}.

\subsection{V851 Cen}
  
An XMM-\emph{Newton} observation of V851 Cen was awarded to us (P.I.
F.~Favata) in AO-1, and was executed in March 2003
(Table~\ref{obslog}).  XMM-\emph{Newton} carries out simultaneous
observation with the EPIC (European Imaging Photon Camera) PN and MOS
detectors (sensitivity range 0.15--15 keV and 0.2--10~keV
respectively), and with the RGS \citep[Reflection Grating
Spectrometer,][]{denher01} ($\lambda\lambda\sim$6--38~\AA,
$\lambda$/$\Delta\lambda\sim$100--500).  The RGS data have been
reduced employing the standard SAS (Science Analysis Software) version
5.4.1 packages, removing in the RGS spectra the time intervals when
the background was higher than 0.4 cts/s in CCD \#9, in order to
ensure a ``clean'' spectrum.  Light curves were obtained by selecting
a circle centered on the source in the EPIC-pn and EPIC-MOS images,
and subtracting the background count rate taken from a nearby area.
The light curve shows a slow decline ($\sim 20$\%) of the source flux
(Fig.~\ref{lcurves}).  High resolution spectra corresponding to the
first and second orders of the RGS have been used to measure the line
fluxes simultaneously, improving the quality of the measurements.
Fluxes measured for second order lines are in good agreement with
those in the first order \citep{sanz03}, hence they can be used to
improve the quality of the measurements due to their better spectral
resolution.

\subsection{$\lambda$~And}   

An observation with the \emph{Chandra} High Energy Transmission
Grating Spectrograph \citep[HETG,][]{wei02} of $\lambda$~And was
retrieved from the \emph{Chandra} archive (Table~\ref{obslog}).  The
HETGS is made of two gratings, HEG (High Energy Grating,
$\lambda\lambda\sim$1.7--15, $\lambda/\Delta\lambda\sim$65--1070),
and MEG (Medium Energy Grating $\lambda\lambda\sim$3--27,
$\lambda/\Delta\lambda\sim$80--970) operating simultaneously.
Standard reduction tasks from the CIAO v2.3 package have been employed
in the reduction of data and the extraction of the HEG and MEG
spectra. The light curve of $\lambda$~And, obtained from the MEG and
HEG first orders spectra, show no significant events, with variations
of less than a 20\% of the total flux (Fig.~\ref{lcurves}).

\subsection{Procyon and $\epsilon$~Eri}   

Observations with the \emph{Chandra} Low Energy Transmission Grating
Spectrograph \citep[LETGS, $\lambda\lambda\sim$3--175,
$\lambda/\Delta\lambda\sim$60-1000][]{wei02} and the High Resolution Camera
(HRC-S) of Procyon and $\epsilon$~Eri were retrieved from the
\emph{Chandra} archive as listed in Table~\ref{obslog}.  Data were
reduced using standard reduction tasks present in the CIAO v2.3
package.  The positive and negative spectral orders were summed for
the fluxes measurement, although at long wavelengths ($\ga 40$~\AA)
separated measurements were made in the plus and minus spectra due to
problems with wavelength calibration or detector gaps. Lines formed in
the first dispersion order, but contaminated with contribution from
higher dispersion orders, were not employed in the analysis. The two
observations available for Procyon were summed in order to achieve
better statistics, resulting in an exposure time of 139~ks.  Finally,
light curves were obtained from the LEG spectra (first and higher
orders) of Procyon and $\epsilon$~Eri as shown in Fig.~\ref{lcurves}.

\begin{figure*}
\begin{center}
\includegraphics[width=0.49\textwidth]{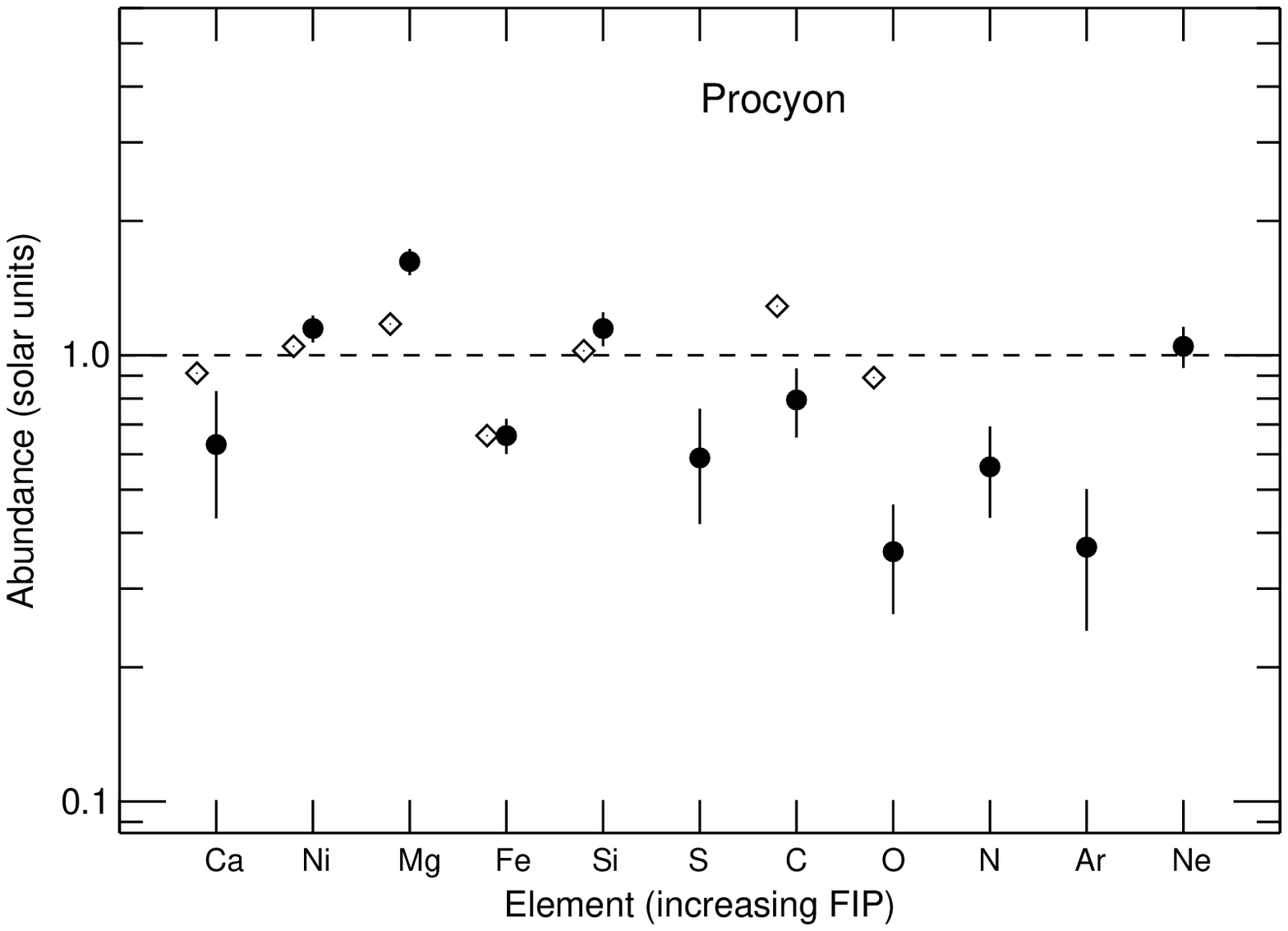}
\includegraphics[width=0.49\textwidth]{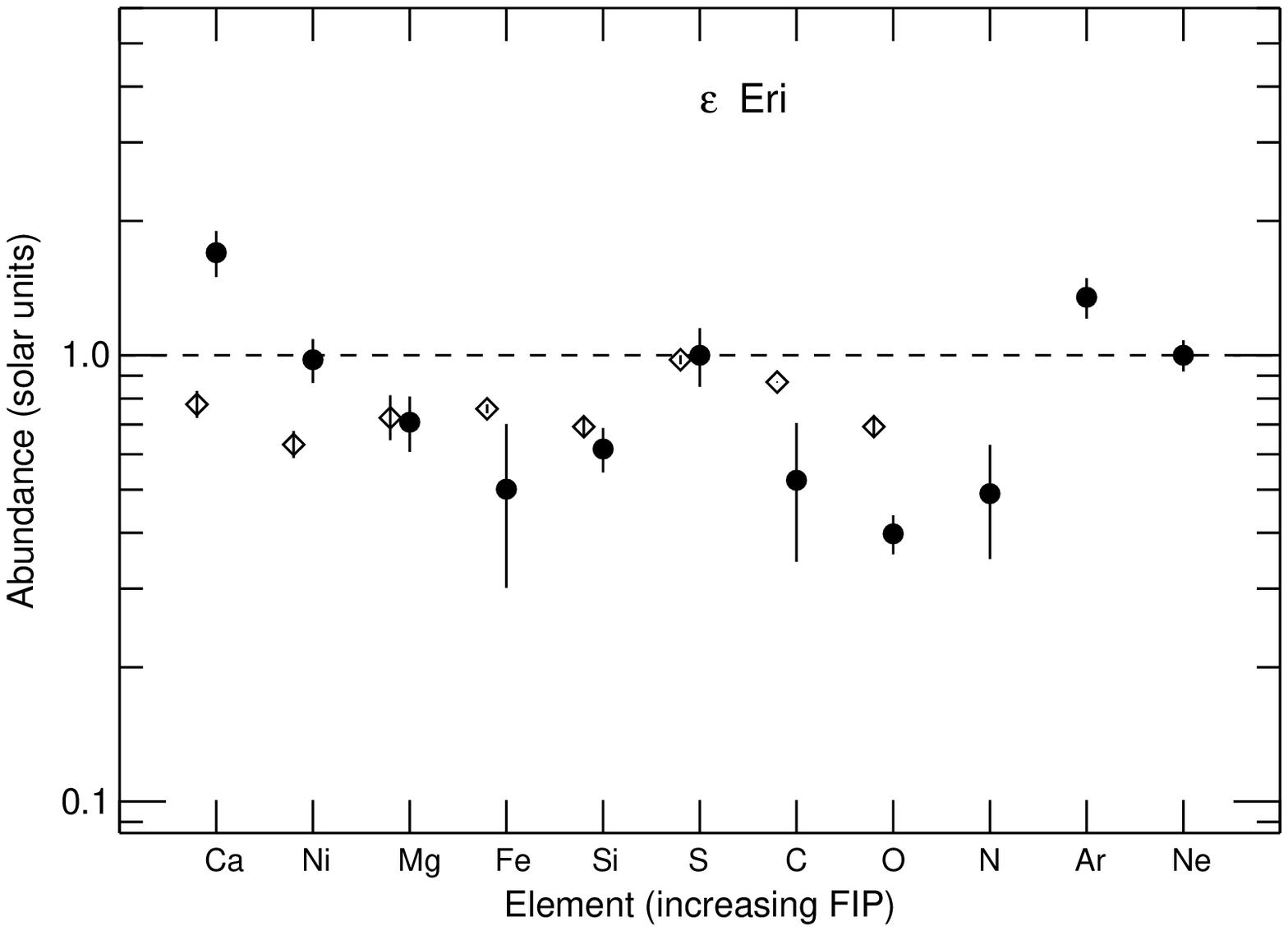}
\includegraphics[width=0.49\textwidth]{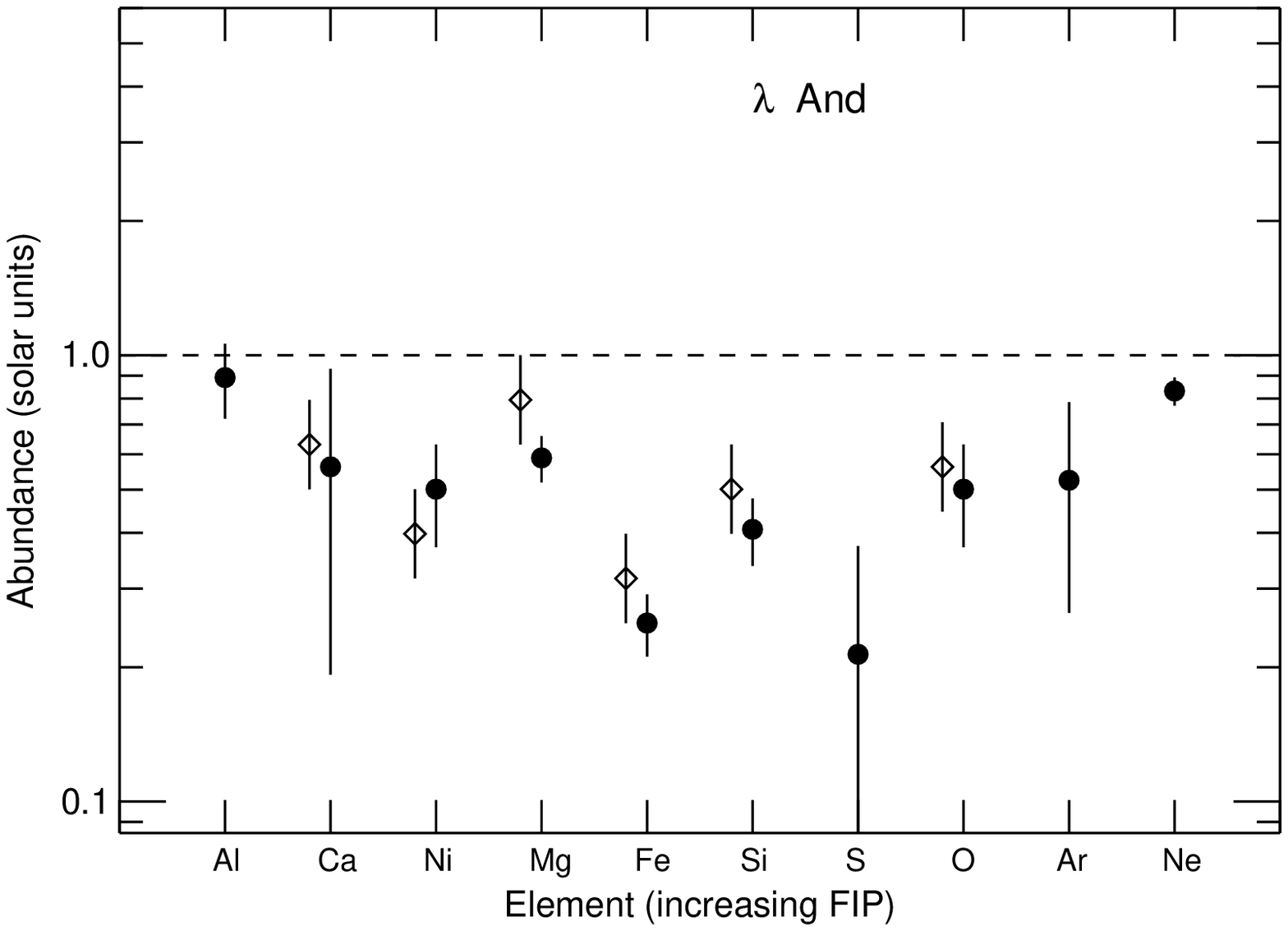}
\includegraphics[width=0.49\textwidth]{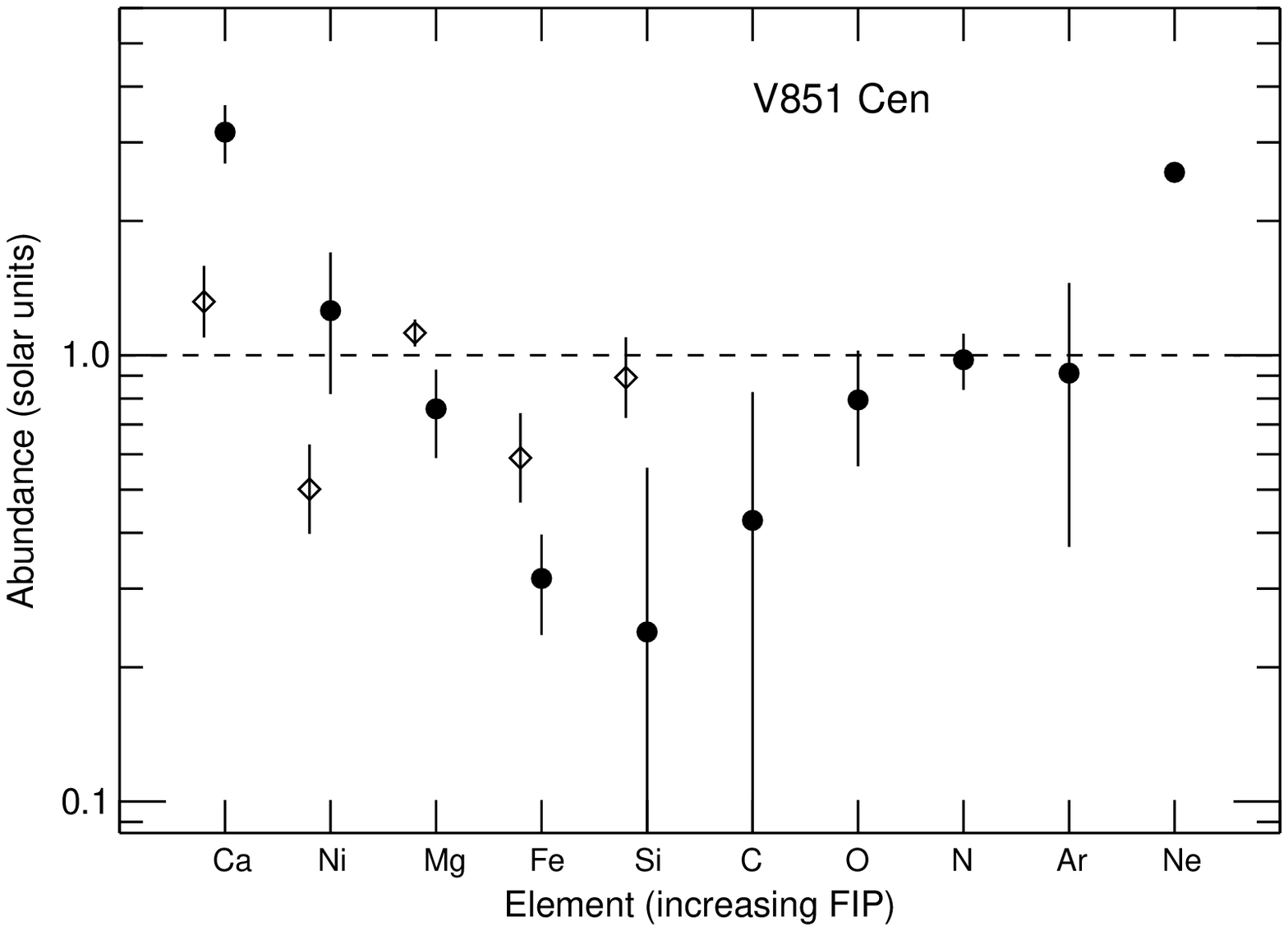}
\end{center}
 \caption{Element abundances in the corona of Procyon,
   $\epsilon$~Eri, $\lambda$~And and V851~Cen (filled circles) with
   respect to solar photospheric values. Open diamonds represent the
   stellar photospheric abundances (see text).  A dashed line
   indicates the adopted solar photospheric abundance \citep{anders}.
   \label{figabund}}
\end{figure*}

\begin{figure*}
\begin{center}
\includegraphics[width=0.49\textwidth]{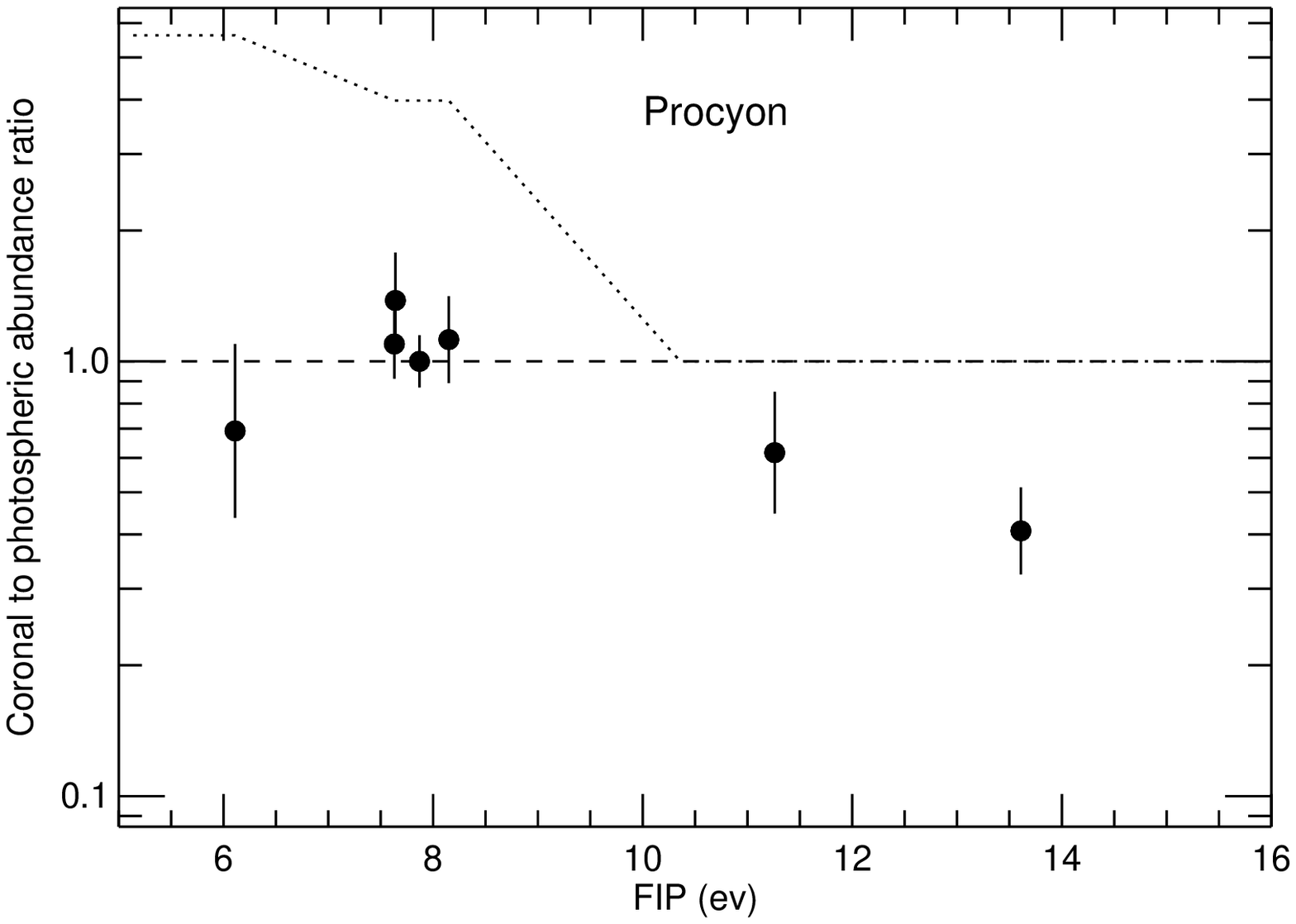}
\includegraphics[width=0.49\textwidth]{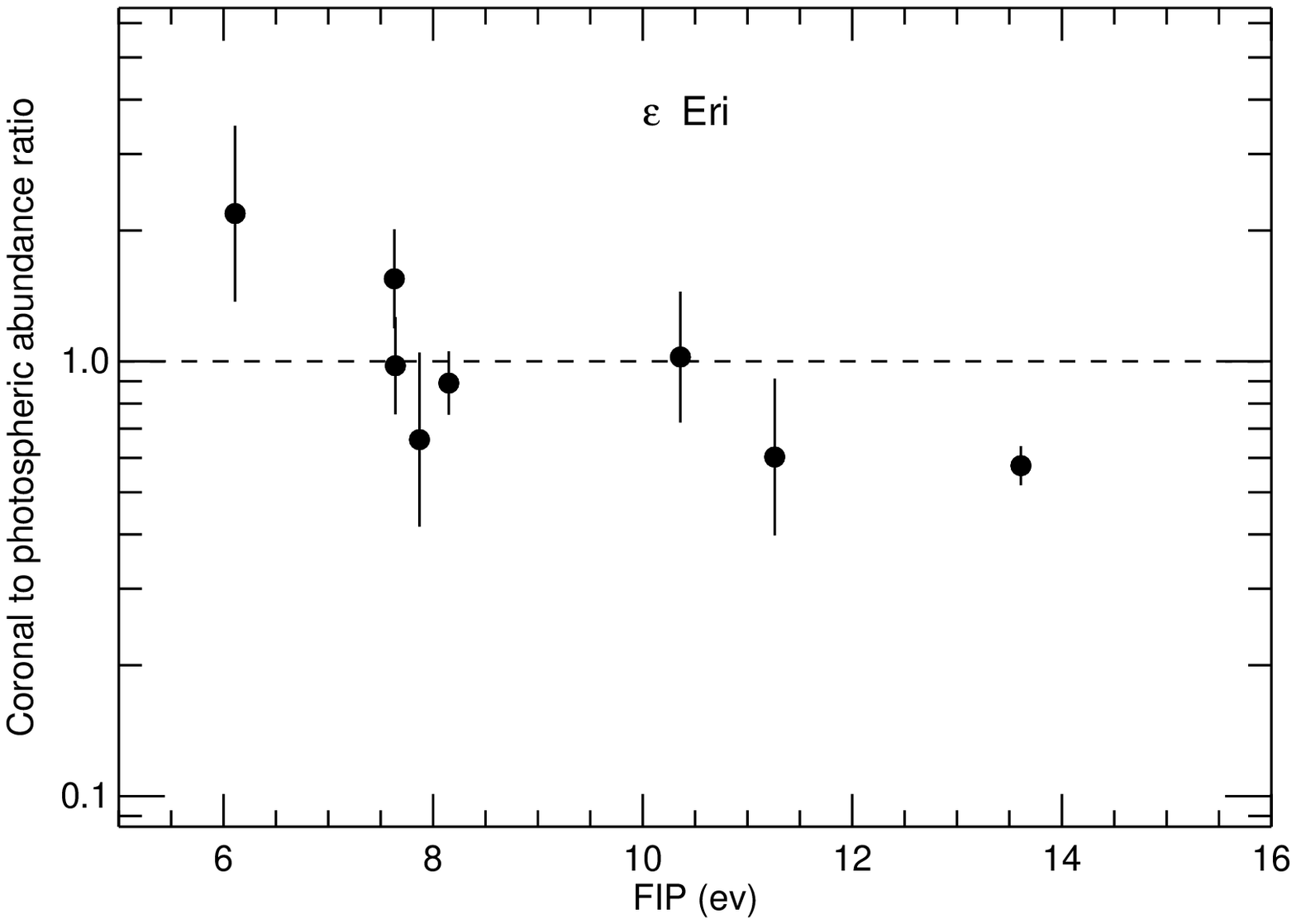}
\includegraphics[width=0.49\textwidth]{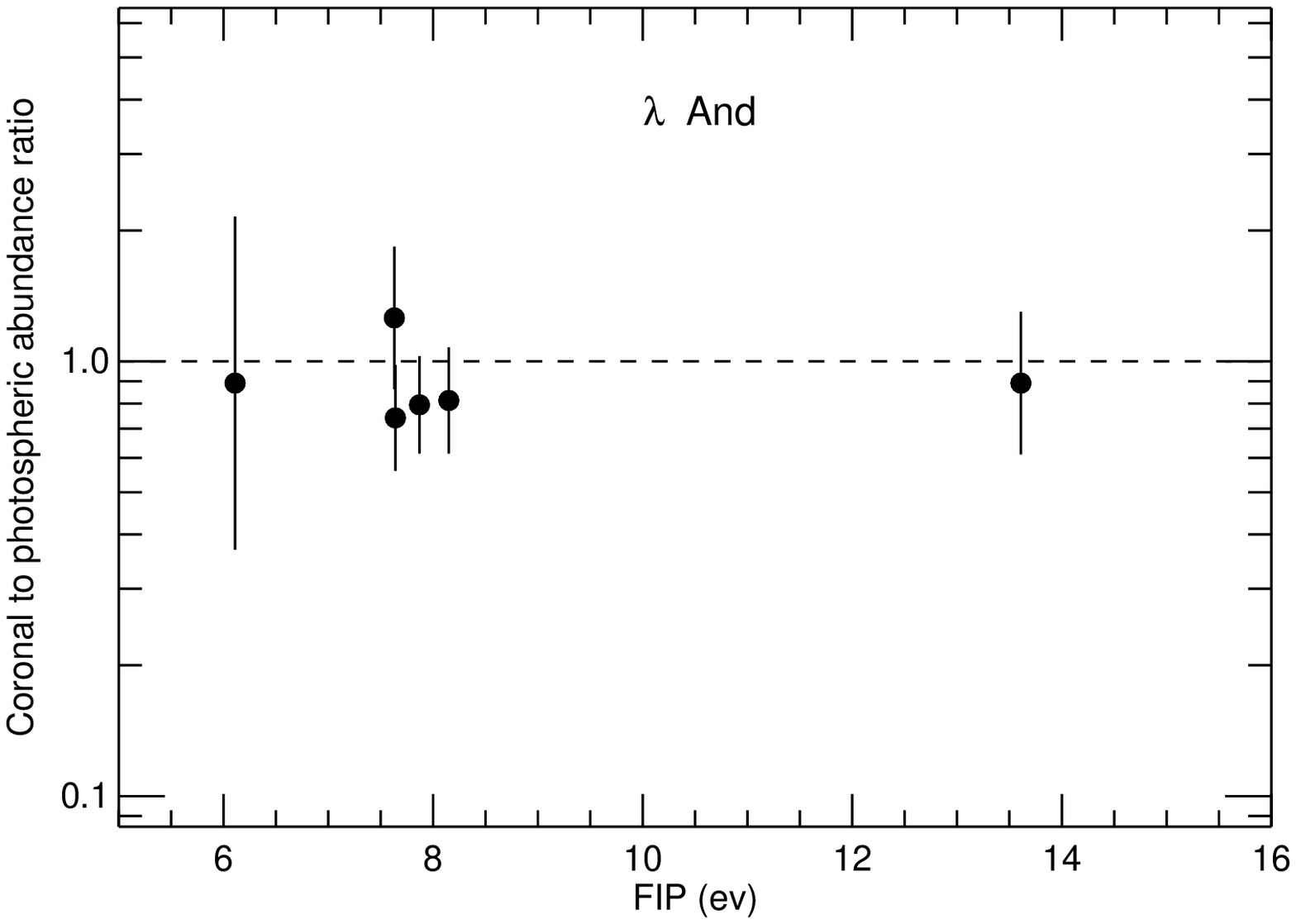}
\includegraphics[width=0.49\textwidth]{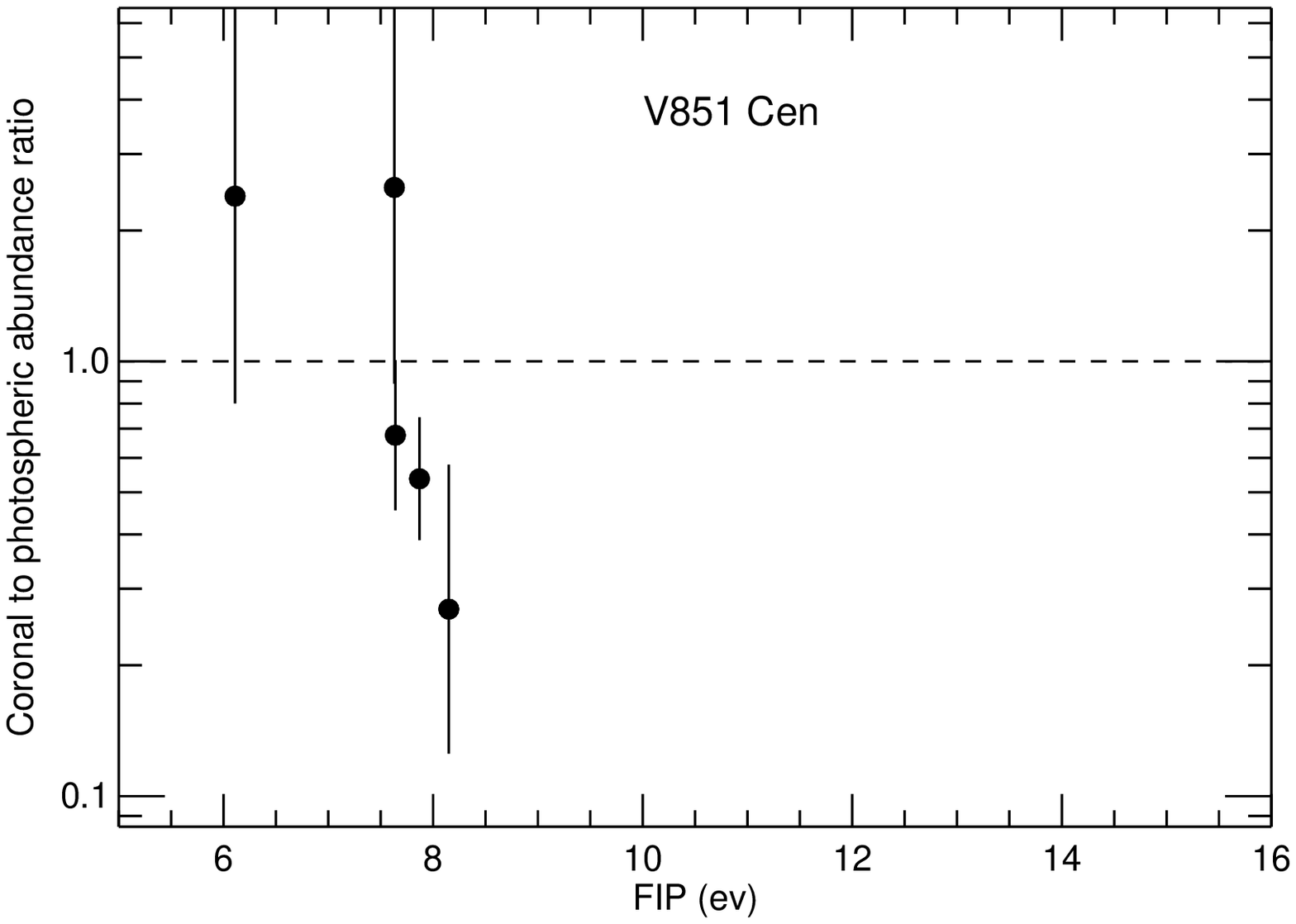}
\end{center}
\caption{Coronal to photospheric abundance ratio for Procyon,
  $\epsilon$~Eri, $\lambda$~And and V851~Cen, with $1\sigma$ error
  bars.  Solar coronal (at $\log T[K]\sim 6.1$) to photospheric
  abundance ratios \citep{feld00} are indicated with a dotted line in
  the Procyon plot.
   \label{figabund2}}
\end{figure*}

\section{Data analysis}\label{sec:analysis}

X-ray spectra from stellar coronae have numerous lines from different
elements formed at different temperatures. This variety of
temperatures in the lines formation makes necessary to determine in
detail the thermal structure of the target star, commonly
parameterized through the emission measure distribution (EMD) as a
function of temperature, defined as $E\!M(T) = \int_{\Delta T} N_H N_e
dV$ [cm$^{-3}$]. In the present work we apply a line-based method in
order to reconstruct the EMD of the stellar coronae \citep[this method
is preferred to the global fitting techniques, see][]{sanz03}, with a
step in temperature of 0.1 dex, the same as employed in the
Astrophysical Plasma Emission Database \citep[APED v1.3,][]{aped},
which is used for comparison of the observed with the predicted line
fluxes as explained below.

\begin{figure*}
\begin{center}
  \includegraphics[width=0.49\textwidth]{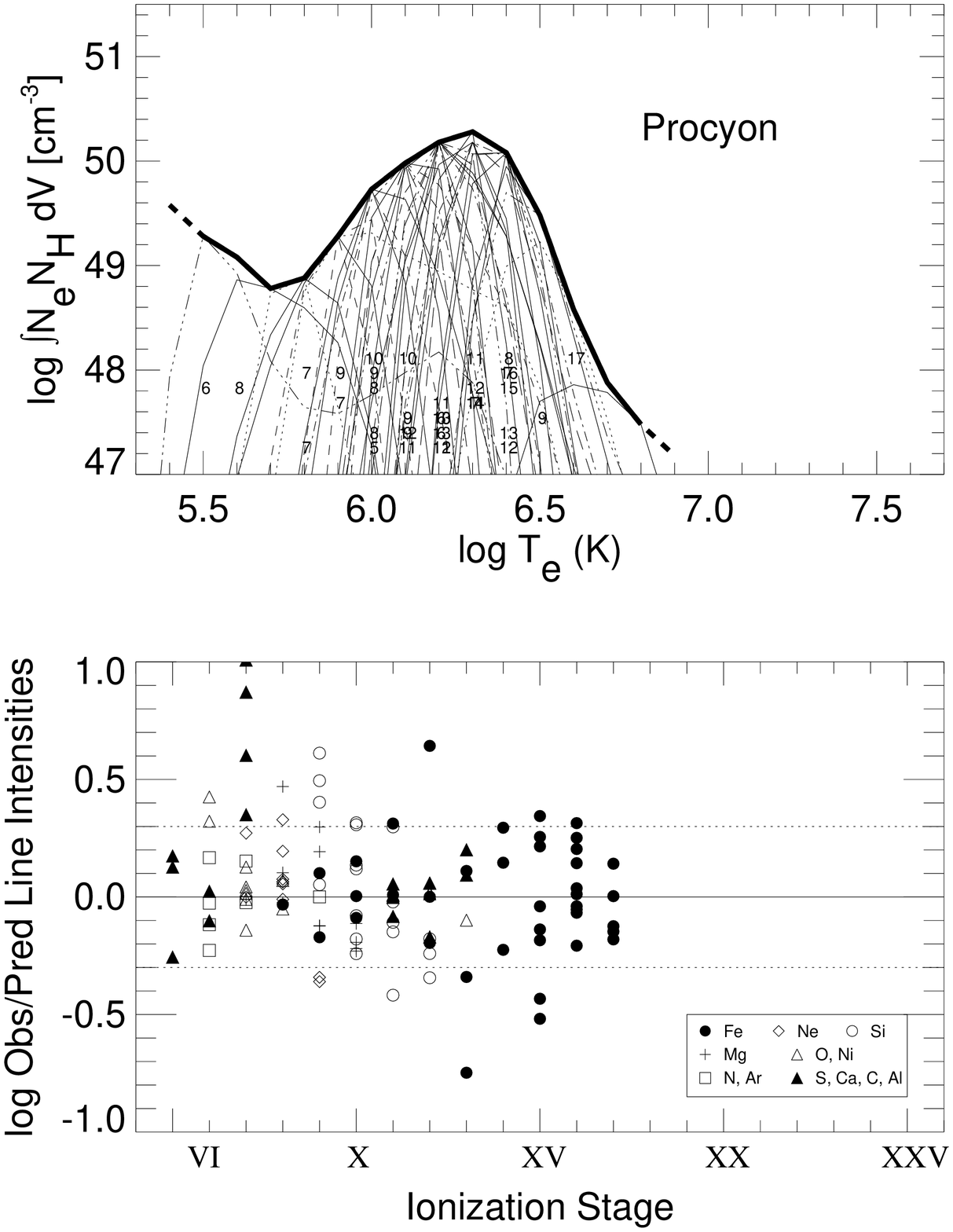}
  \includegraphics[width=0.49\textwidth]{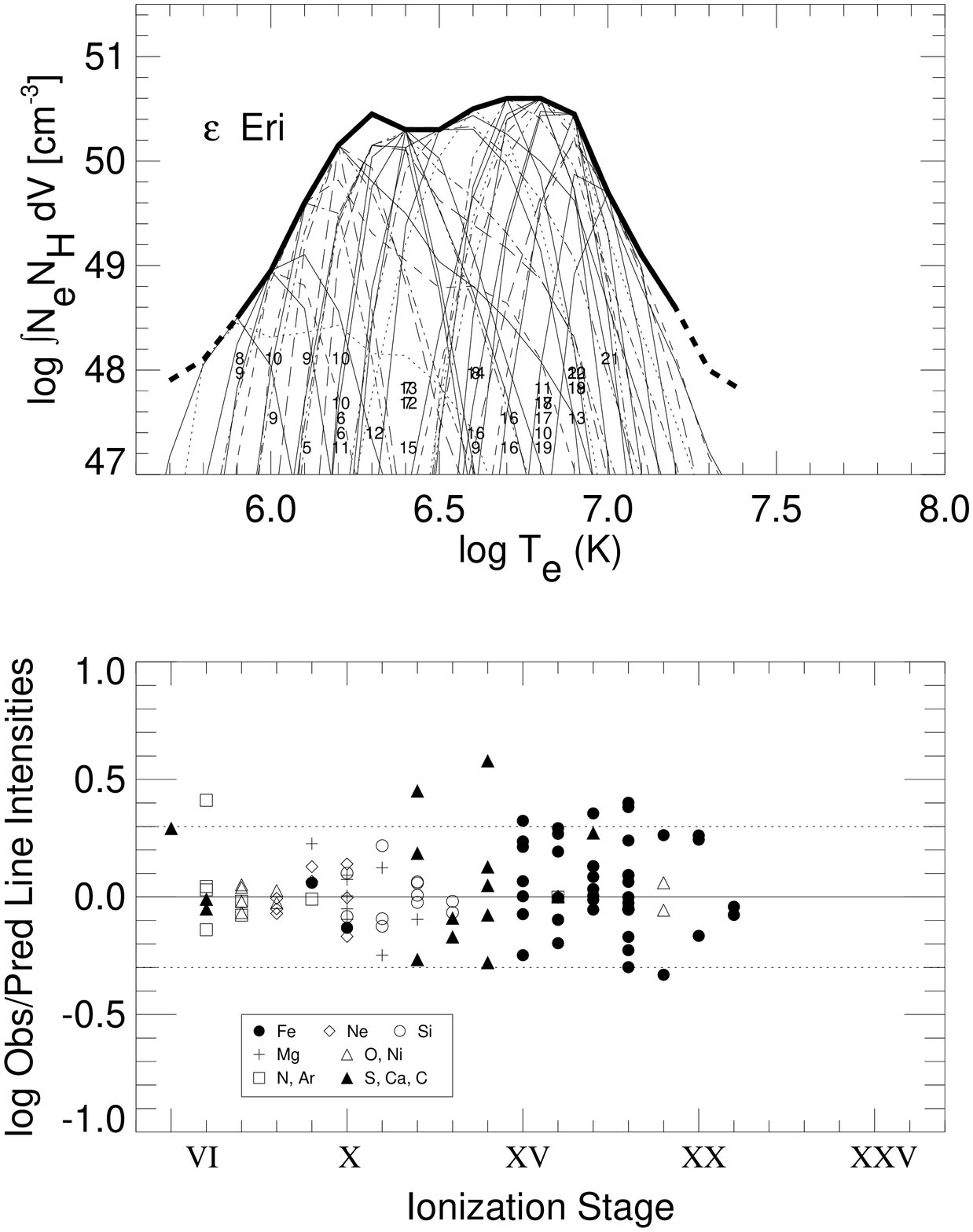}
  \includegraphics[width=0.49\textwidth]{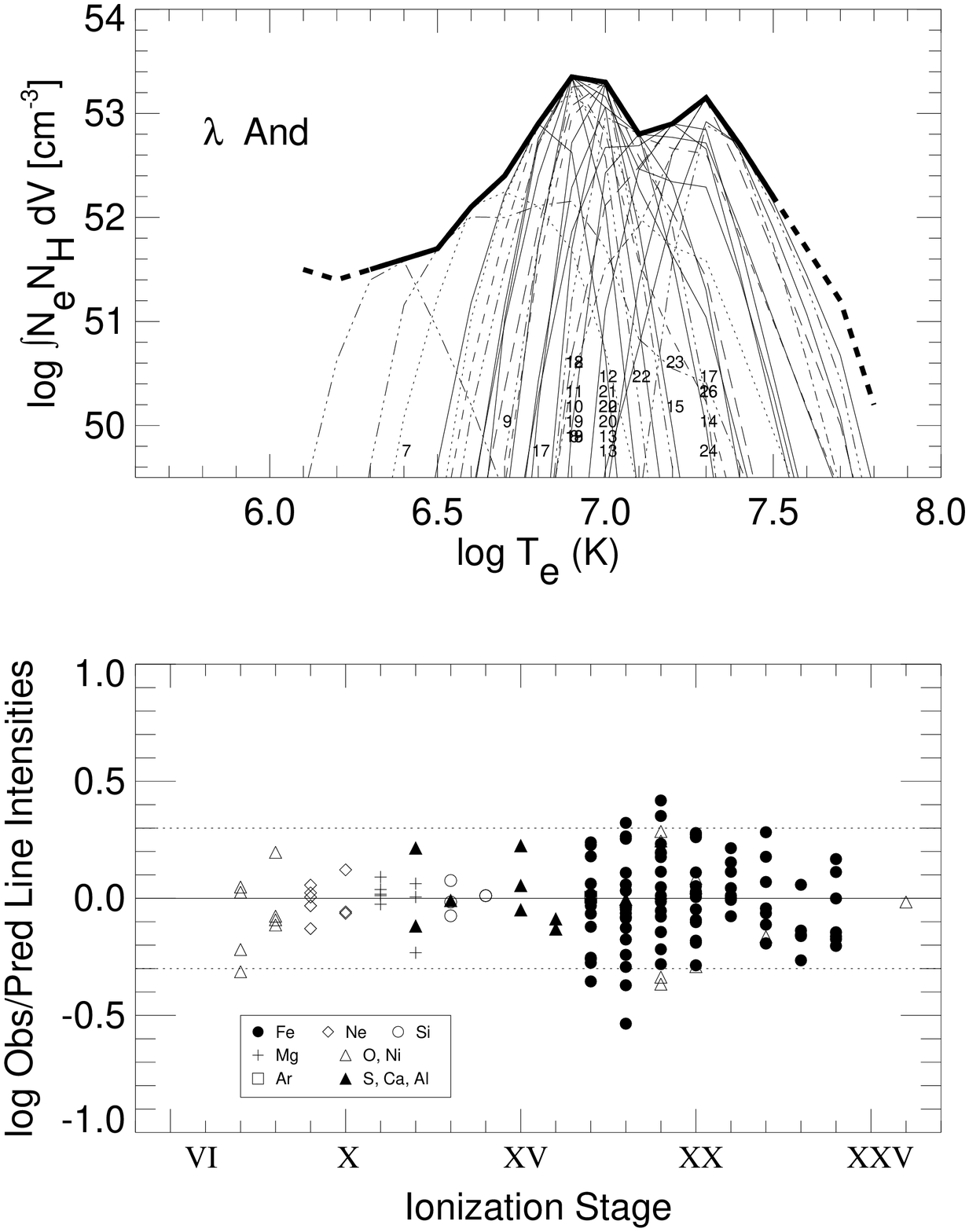}
  \includegraphics[width=0.49\textwidth]{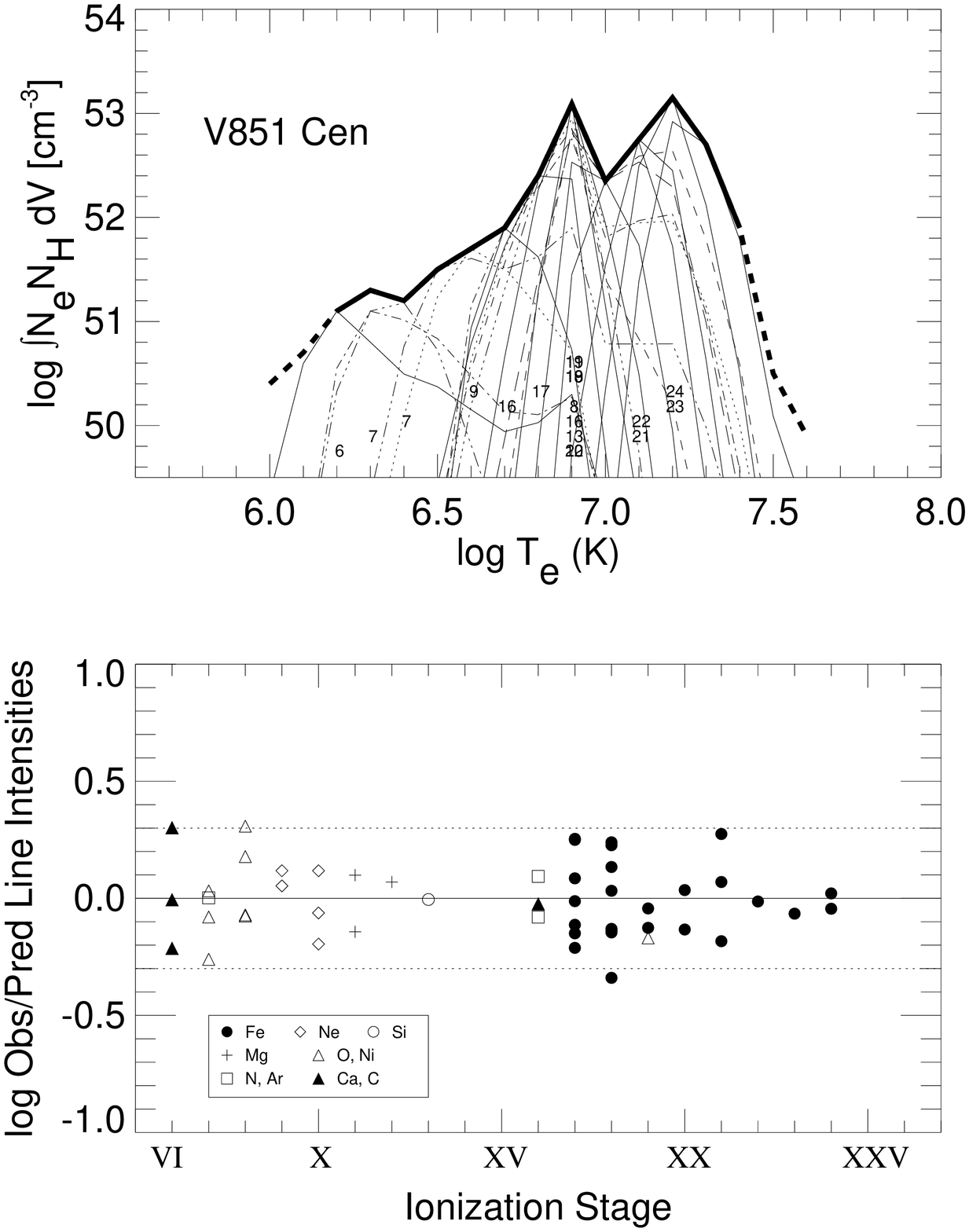}
\end{center}
 \caption{Emission Measure Distribution for Procyon, $\epsilon$~Eri,
   $\lambda$~And and V851~Cen.  Thin lines represent the relative
   contribution function for each ion (the emissivity function
   multiplied by the EMD at each point). Small numbers indicate the
   ionization stages of the species. Also plotted are the observed to
   predicted line flux ratios for the ion stages in the upper figure.
   The dotted lines denote a factor of 2.\label{emds}}
\end{figure*}

The EMD reconstruction was conducted by measuring the fluxes from
spectral lines present in the high resolution X-ray spectra, following
\citet{sanz03}.  The {\em Interactive Spectral Interpretation System}
\citep[ISIS,][]{isis} software package, provided by the MIT/CXC, was
employed to measure the line fluxes, through convolution of the
spectral response. These measured line fluxes were then corrected from
interstellar medium absorption (ISM) (although such correction is
important only for very few lines).
Theoretical fluxes can be predicted by assuming an initial EMD which
is combined with the emissivity functions of the different spectral
lines. These fluxes are then compared with the observed fluxes, and
the EMD is changed to determine a solution that shows a better
agreement between predicted and observed fluxes. An iterative process,
as explained in \citet{sanz03} results in the ``best'' solution for
both the EMD and the abundance pattern. Although this solution is not
unique, the use of a large number of spectral lines with a good
temperature coverage constrains the interval of possible solutions to
a small range of values around the solution proposed.  Error bars to
the EMD can be calculated by means of a Montecarlo method, consisting
on the variation of the observed fluxes by $\pm 1 \sigma$ and the
determination of the best solution among 1000 possible variations
around the initial EMD solution. Such exercise is made for 1000 sets
of possible line fluxes, providing error bars for the solution.  The
abundances of the different elements are determined during the
calculation of the EMD, with error bars resulting from the quadratic
summation of the errors on the measurement of observed fluxes ($\Delta
F_{\rm obs}/F_{\rm obs})$, and the dispersion of the
predicted-to-observed line ratios [$(F_{\rm obs}-F_{\rm pred})/F_{\rm
  obs}$] of lines of the same element (the latter permits to evaluate
indirectly the errors related to atomic models).  The abundances of
the elements are determined {\em relative} to Fe, with the [Fe/H]
abundance derived afterwards from the best fit to the global spectrum
for the given EMD and abundances pattern. The coronal abundances
determined, with their photospheric counterparts are shown in
Figs.~\ref{figabund} and~\ref{figabund2}, and are listed in
Table~\ref{tab:abundances}. EMD distributions (Table~\ref{tab:emds})
are shown in Fig.~\ref{emds}. The line fluxes measured in the four
targets are listed in
Tables~\ref{tab:flprocyon}--\ref{tab:flv851}\footnote{These tables are
available in electronic format only.}

\section{Results}\label{sec:results}

\subsection{Procyon}\label{sec:procyon}

Procyon ($\alpha$ CMi A, HR 2943, HD 61421) is a well studied nearby
F4IV star \citep[3.497~pc][]{hippa}. A rotational period of $\sim
9.1$~d can be deduced from $v\,{\rm sin}\,i=6.1$ km/s \citep{med99}
and $i=32$\degr \citep{irw92}.  Photospheric abundances on different
elements have been determined by many authors, resulting in a
metallicity close to solar.  In this work we will use the values
calculated by \citet{edv93}, with C abundance from \citet{var99}.
First studies on the coronal abundances were carried out using EUVE
observations: \citet{dra95} calculated, using line-based analysis, the
EMD for low and high FIP elements separately, but they did not find
any ``FIP'' or ``inverse FIP'' effect. Global fits to
\emph{Chandra}/LETG and XMM-\emph{Newton} high-resolution spectra were
employed by \citet{raa02} to determine the abundances of Procyon, also
concluding that no FIP-related effects are present in the corona of
Procyon (although no comparison with photospheric values is present in
their study).  

In the present work line fluxes were corrected from the ISM assuming a
value of $\log~N_{\rm H}({\rm cm}^{-2})\sim 18.2$ \citep{paper3}.
Several problems affect the analysis of the \emph{Chandra} Procyon
spectra.  First, the shape of the continuum of a relatively cool
corona like that of Procyon, in the wavelength range covered here,
does not allow to establish the [Fe/H] coronal abundance accurately (a
wide range of values show similar line-to-continuum ratios), thus we
fixed it at its photospheric value (note that the rest of coronal
abundances are calculated relative to Fe, and the general level of the
EMD depends on [Fe/H] too). Also, the determination of coronal
abundances is affected, in the LETG wavelenght range, by the lack of a
complete temperature coverage provided by Fe (only two lines are
formed at $\log~T[K]\la 6.2$), while many lines from other elements
have little temperature overlap with Fe lines.  However it is possible
to include in the analysis the Fe lines observed with EUVE
(\ion{Fe}{xi}--{\scshape xvi}), obtaining a better temperature
coverage for one element, useful for the determination of abundances
of other elements.  Measurements of EUVE Fe line fluxes in the range
170--370~\AA\ were made by \citet{paper3} from observations taken in
1993, 1994 and 1999 (the latter is partially simultaneous with the
LETG campaign, and no substantial changes were detected in the EUV
flux level). The inclusion of the EUVE lines allows to cover the range
$\log~T(K)\sim 5.6$--6.8, where almost all lines of other elements are
also formed.

\begin{table*}
\caption{Coronal and photospheric abundances of the elements ([X/H], solar
  units) in the target stars.}\label{tab:abundances} 
\tabcolsep 3.pt
\begin{center}
\begin{footnotesize}
 \begin{tabular}{lrrrrrrrrrr}
\hline \hline
{X} & {FIP} & Reference$^a$ & \multicolumn{2}{c}{Procyon} &
\multicolumn{2}{c}{$\epsilon$~Eri} & \multicolumn{2}{c}{$\lambda$~And}
& \multicolumn{2}{c}{V851 Cen}  \\ 
    &  eV   & solar value & Photosphere & Corona$^b$ & Photosphere & Corona &
Photosphere & Corona &  Photosphere & Corona \\
\hline
 Al &  5.98 & 6.47 & 0.00 &  0.45$\pm$ 0.12 & $-$0.12$\pm$ 0.02 & \ldots & \ldots & $-$0.05$\pm$ 0.17 &  0.25$\pm$ 0.05 & \ldots \\
 Ca &  6.11 & 6.36 & $-$0.04 & $-$0.20$\pm$ 0.20 & $-$0.11$\pm$ 0.03 &  0.23$\pm$ 0.20 & $-$0.20$\pm$ 0.10 & $-$0.25$\pm$ 0.37 &  0.12$\pm$ 0.08 &  0.50$\pm$ 0.47 \\
 Ni &  7.63 & 6.25 & 0.02 &  0.06$\pm$ 0.08 & $-$0.20$\pm$ 0.03 & $-$0.01$\pm$ 0.11 & $-$0.40$\pm$ 0.10 & $-$0.30$\pm$ 0.13 & $-$0.30$\pm$ 0.10 &  0.10$\pm$ 0.44 \\
 Mg &  7.64 & 7.58 & 0.07 &  0.21$\pm$ 0.11 & $-$0.14$\pm$ 0.05 & $-$0.15$\pm$ 0.10 & $-$0.10$\pm$ 0.10 & $-$0.23$\pm$ 0.07 &  0.05$\pm$ 0.03 & $-$0.12$\pm$ 0.17 \\
 Fe &  7.87 & 7.50 & $-$0.18 & $-$0.18$\pm$ 0.06 & $-$0.12$\pm$ 0.01 & $-$0.30$\pm$ 0.20 & $-$0.50$\pm$ 0.10 & $-$0.60$\pm$ 0.05 & $-$0.23$\pm$ 0.10 & $-$0.50$\pm$ 0.10 \\
 Si &  8.15 & 7.55 & 0.01 &  0.06$\pm$ 0.10 & $-$0.16$\pm$ 0.02 & $-$0.21$\pm$ 0.07 & $-$0.30$\pm$ 0.10 & $-$0.39$\pm$ 0.07 & $-$0.05$\pm$ 0.09 & $-$0.62$\pm$ 0.32 \\
  S & 10.36 & 7.21 & \ldots & $-$0.23$\pm$ 0.17 & $-$0.01$\pm$ 0.01 &  0.00$\pm$ 0.15 & \ldots & $-$0.67$\pm$ 0.16 & \ldots & $-$1.05$\pm$ 1.33 \\
  C & 11.26 & 8.56 & 0.11 & $-$0.10$\pm$ 0.14 & $-$0.06\hfill\hspace{1ex} & $-$0.28$\pm$ 0.18 & \ldots & \ldots & \ldots & $-$0.37$\pm$ 0.40 \\
  O & 13.61 & 8.93 & $-$0.05 & $-$0.44$\pm$ 0.10 & $-$0.16$\pm$ 0.02 & $-$0.40$\pm$ 0.04 & $-$0.25$\pm$ 0.10 & $-$0.30$\pm$ 0.13 & \ldots & $-$0.09$\pm$ 0.23 \\
  N & 14.53 & 8.05 & \ldots & $-$0.25$\pm$ 0.13 & \ldots & $-$0.31$\pm$ 0.14 &
 \ldots & \ldots & \ldots & 0.00$\pm$ 0.14 \\
 Ar & 15.76 & 6.56 & \ldots & $-$0.43$\pm$ 0.13 & \ldots &  0.13$\pm$ 0.14 & \ldots & $-$0.28$\pm$ 0.26 & \ldots & $-$0.04$\pm$ 0.54 \\
 Ne & 21.56 & 8.09 & \ldots &  0.02$\pm$ 0.11 & \ldots &  0.00$\pm$ 0.08 & \ldots & $-$0.08$\pm$ 0.06 & \ldots &  0.41$\pm$ 0.13 \\
\hline
\end{tabular}
\end{footnotesize}
\end{center}
$^a$ Solar photospheric abundances from \citet{anders}, adopted in
this work, are expressed in logarithmic scale. 
Note that several values have been
updated in the literature, most notably the cases of O \citep[now
  $\sim$8.7,][]{all01,hol01} and C \citep[now 8.39,][]{all02}.

$^b$ The [Fe/H] coronal abundance of Procyon is set to its
photospheric value, with nominal error bars determined from the
observed-to-predicted fluxes dispersion. 
\end{table*}

The abundances determined for Procyon do not show substantial
differences between photospheric and coronal relative abundances (note
that no error bars are published for the photospheric abundances),
with the most notable difference being a relative underabundance of O
in the corona. Nevertheless, an interesting feature arises when
fitting the EMD and abundances. Assuming that the EMD (see
Fig.~\ref{emds}) is correct for all elements, a higher (relative to
Fe) abundance of S, O, Ne, Si and Mg at lower temperatures would
improve the fit of their respective lines (the case of the
\ion{S}{vii} lines is very remarkable), while elements like Ni, Ca,
Al, C or N cover a gradient of temperatures too small to notice any
trend. This effect could be also caused by the Fe
abundance increasing 
with temperature, from the minimum of the EMD at $\log T(K)\sim 5.8$ to
the maximum at $\log T(K) \sim 6.3$. This might indicate a trend of a
mild FIP effect which increases with temperature in the corona of
Procyon \citep[][also discuss a possible temperature dependence to
    the coronal abundances in Procyon]{dra95}.  
In the solar corona, the FIP effect is indeed larger at
$\log T(K) \sim 6.15$ than at $\log T(K) \la 5.9$ \citep{feld00}.
Although suggestive, this hypothesis cannot presently be confirmed
with high confidence for Procyon given the present uncertainties in
atomic models and in the determination of the EMD of this star.
Similarly, an increase in the [Ne/Fe] abundance with temperature
between $\log T(K) \sim 5.8$ and $\log T(K) \sim 6.6$ would explain
the discrepancies found in the determination of the coronal Ne
abundance of AD Leo from \ion{Ne}{viii} lines \citep{paperadleo} and
from \ion{Ne}{ix} and \ion{Ne}{x} lines \citep{mag02}.

\subsection{$\epsilon$ Eri}

As one of the nearest cool stars ($d = 3.22$~pc), $\epsilon$~Eri (HD
22049, HR 1084) is frequently studied regarding its activity,
abundances, and more recently, stellar planetary companions \citep[see
e.g.,][]{hat00}. With $P_{\rm phot}\sim 11.3$~d \citep{bal83},
$\epsilon$~Eri (K2V) is a star with an intermediate activity level.
Photospheric abundances derived by \citet{zhao00} are used here for
comparison with the coronal values, and a value of $\log N_{\rm H}
({\rm cm}^{-2})\sim 18.1$ \citep{paper3} is adopted to correct for
ISM absorption.  The first high-resolution analysis on coronal
abundances of $\epsilon$~Eri was made by \citet{lam96} who analyzed
EUVE spectra, calculating separate line-based EMDs for low- and
high-FIP elements, concluding that a solar-like FIP effect was present
in the corona of $\epsilon$~Eri. Coronal electron density diagnostics
were reported by \citet{ness02} from the \emph{Chandra}/LETG
observations of $\epsilon$~Eri. 

In the present work we derive the coronal abundance of 11 elements
from the \emph{Chandra}/LETG spectrum of $\epsilon$~Eri. The corona is
dominated by emitting material in the range $\log T(K)\sim 5.8$--7.2,
with a distribution compatible with the presence of a peak at $\log
T(K)\sim 6.5$ as EUVE data suggest \citep{paper3}, but also with two
peaks at $\log T(K) \sim 6.3$ and 6.8, as the present data, also
compatible with EUVE fluxes, suggest. The [Fe/H] coronal abundance is
better constrained than for Procyon, with a 0.2~dex uncertainty; the
agreement between predicted and observed fluxes is also better than
for Procyon.  A mild FIP-like bias is detected in the corona of
$\epsilon$~Eri, affecting only Ca and Ni (Fig.~\ref{figabund2}), with
a progressive lower coronal abundance for elements with higher FIP --
while in the Sun the FIP effect consists in the enhancement by more
than a factor of 3 of the elements with ${\rm FIP} \la 10$~eV.

\subsection{$\lambda$~And}

$\lambda$~And (HD 222107, HR 8961) is an active binary system with a
G8~IV-III primary and an unseen companion in an orbit with $P_{\rm
  orb}=20.5212$~d \citep{walker}. $\lambda$~And is one of the few
active stars for which measurements of photospheric abundances for a
good number of elements \citep{sav94,don95} are available, also thanks
to its low rotation rate \citep[$P_{\rm phot}=54.33$~d,][]{hippa}.
Coronal abundances from high resolution spectra were not obtained from
EUVE data because the ISM absorption did not allow to measure fluxes
of lines for elements other than Fe \citep[although the spectrum is
suggestive of a low value of {[Fe/H]}, see][]{paperland}, and the
global fit of the XMM/RGS spectra gave no clear trend with FIP
\citep{aud03}.  A preliminary analysis of the \emph{Chandra}/HETG and
LETG spectra was reported by \citet{dup00} and \citet{bri03}
respectively.

\begin{table}
\caption{Emission Measure Distribution of the target stars}\label{tab:emds}
\tabcolsep 3.pt
\begin{center}
\begin{small}
\begin{tabular}{lrrrr}
\hline \hline
{log~$T$} & \multicolumn{4}{c}{log $\int N_{\rm e} N_{\rm H} {\rm d}V$
  (cm$^{-3}$)$^a$} \\
(K) & {Procyon} & {$\epsilon$~Eri} & {$\lambda$~And} & {V851 Cen}  \\ 
\hline
5.4 & 49.58\hfill\hspace{1ex}  & \ldots  & \ldots   & \ldots   \\
5.5 & 49.28$^{+0.55}_{-0.05}$  & \ldots  & \ldots  & \ldots  \\
5.6 & 49.08$^{+0.40}_{-0.30}$  & \ldots  & \ldots  & \ldots  \\
5.7 & 48.78$^{+0.20}_{-0.40}$  & 47.60\hfill\hspace{1ex}  & \ldots  & \ldots  \\
5.8 & 48.88$^{+0.20}_{-0.40}$  & 47.80\hfill\hspace{1ex}  & \ldots  & \ldots  \\
5.9 & 49.28$^{+0.20}_{-0.20}$  & 48.20$^{+0.20}_{-0.30}$  & \ldots  & \ldots  \\
6.0 & 49.73$^{+0.05}_{-0.40}$  & 48.65$^{+0.20}_{-0.30}$  & \ldots  & 50.40\hfill\hspace{1ex}  \\
6.1 & 49.98$^{+0.40}_{-0.05}$  & 49.30$^{+0.20}_{-0.20}$  & 50.90\hfill\hspace{1ex}  & 50.70\hfill\hspace{1ex}  \\
6.2 & 50.18$^{+0.05}_{-0.20}$  & 49.85$^{+0.10}_{-0.20}$  & 50.80\hfill\hspace{1ex}  & 51.10$^{+0.05}_{-0.40}$  \\
6.3 & 50.28$^{+0.05}_{-0.10}$  & 50.15$^{+0.10}_{-0.20}$  & 50.90$^{+0.20}_{-0.30}$  & 51.30$^{+0.05}_{-0.35}$  \\
6.4 & 50.08$^{+0.10}_{-0.05}$  & 50.00$^{+0.20}_{-0.10}$  & 51.00$^{+0.20}_{-0.30}$  & 51.20$^{+0.05}_{-0.35}$  \\
6.5 & 49.48$^{+0.05}_{-0.30}$  & 50.00$^{+0.20}_{-0.30}$  & 51.10$^{+0.20}_{-0.30}$  & 51.50$^{+0.10}_{-0.40}$  \\
6.6 & 48.58$^{+0.20}_{-0.40}$  & 50.20$^{+0.20}_{-0.20}$  & 51.50$^{+0.10}_{-0.40}$  & 51.70$^{+0.20}_{-0.40}$  \\
6.7 & 47.88$^{+0.30}_{-0.40}$  & 50.30$^{+0.15}_{-0.25}$  & 51.80$^{+0.05}_{-0.35}$  & 51.90$^{+0.05}_{-0.35}$  \\
6.8 & 47.48$^{+0.30}_{-0.40}$  & 50.30$^{+0.05}_{-0.35}$  & 52.30$^{+0.05}_{-0.15}$  & 52.40$^{+0.05}_{-0.30}$  \\
6.9 & 47.18\hfill\hspace{1ex}                   & 50.15$^{+0.15}_{-0.05}$  & 52.75$^{+0.05}_{-0.05}$  & 53.09$^{+0.05}_{-0.05}$  \\
7.0 &  \ldots                  & 49.40$^{+0.40}_{-0.10}$  & 52.70$^{+0.15}_{-0.05}$  & 52.35$^{+0.20}_{-0.30}$  \\
7.1 &  \ldots                  & 48.80$^{+0.40}_{-0.30}$  & 52.20$^{+0.05}_{-0.20}$  & 52.75$^{+0.20}_{-0.20}$  \\
7.2 &  \ldots                  & 48.30$^{+0.30}_{-0.30}$  & 52.30$^{+0.05}_{-0.40}$  & 53.15$^{+0.05}_{-0.30}$  \\
7.3 &  \ldots                  & 47.70\hfill\hspace{1ex}                   & 52.55$^{+0.05}_{-0.40}$  & 52.70$^{+0.30}_{-0.20}$  \\
7.4 &  \ldots                  & 47.50\hfill\hspace{1ex}                   & 52.10$^{+0.20}_{-0.30}$  & 51.90$^{+0.40}_{-0.40}$  \\
7.5 &  \ldots                  &  \ldots                  & 51.60$^{+0.40}_{-0.30}$  & 50.50\hfill\hspace{1ex}  \\
7.6 &  \ldots                  &  \ldots                  & 51.10\hfill\hspace{1ex}  & 50.50\hfill\hspace{1ex}  \\
7.7 &  \ldots                  &  \ldots                  & 50.60\hfill\hspace{1ex}  & \ldots  \\
7.8 &  \ldots                  &  \ldots                  & 49.60\hfill\hspace{1ex}  & \ldots  \\
\hline
\end{tabular}
\end{small}
\end{center}
$^a$Emission Measure, where $N_{\rm e}$ 
and $N_{\rm H}$ are electron and hydrogen densities, in
cm$^{-3}$. Error bars provided are not independent
between the different temperatures, see text.
\end{table}

The low ISM absorption\footnote{Note that this value of the ISM
  absorption, combined with the distance to the star, is too
  high to see lines from elements other than Fe that were measured for
  other stars (usually at $\lambda \ga$200~\AA) with EUVE, 
  but it is a low value to notice its effects in the HETG spectral
  range.} 
\citep[\protect{$\log N_{\rm H}[{\rm cm}^{-2}]
  = 18.45$},][]{wood96} makes the correction of fluxes unnecessary
(affecting the line fluxes at less than the 1\% level).  The high
spectral resolution and good statistics achieved in the observation
allow to obtain the best determination of the element abundances of
the four stars, also supported by a good agreement between
observed and predicted fluxes for most lines. The fluxes measured in
the EUV range \citep{paperland} are also well predicted using the EMD
determined here. The coronal abundances determined here are remarkably
consistent with the photospheric values calculated by \citet{don95},
and they are also close to the values calculated by \citet{sav94}.

\subsection{V851 Cen}

V851 Cen (HD 119285) is an active binary system with a K2IV-III
primary and an unseen companion, with a photometric period of 12.05~d
\citep{llo87,cuti01}, slightly longer than its orbital period
\citep[$P_{\rm orb}=11.989$ d,][]{saa90}.  V851 Cen has a small
projected rotational velocity \citep[$v \sin i=6.5$ \kms,][]{saa90}
which allowed \citet{katz03} and \citet{morel03} to derive accurate
photospheric abundances, revising the value of [Fe/H]$\sim -0.6$
determined by \citet{ran93} to a value of [Fe/H]$\sim -0.2$.  In the
present work the coronal abundances and thermal structure of this star
is determined for the first time. A correction of the observed fluxes
due to the absorption of the ISM was applied, assuming $\log N_{\rm
  H}({\rm cm}^{-2})\sim 19.5$, the value determined for the nearby
star $\beta$~Cen \citep{fru94}\footnote{Uncertainties in the
  determination of the ISM absorption are of minor importance given
  the small correction (at the $\sim2$\% level at 20~\AA) to be
  applied to the line fluxes.}. We use the photospheric abundances
provided by the ``Method 1'' of \citet{morel03}, as shown in
Fig.~\ref{figabund}. When the coronal abundances are compared to solar
photospheric ones, V851 Cen shows a behavior very similar to that of
stars like AR Lac or AB Dor, with a decrease of abundance with FIP
until Si, and a progressive increment for elements with high FIP.
However, when these abundances are compared with the stellar
photospheric values, the conclusions are different, with only a small
depletion of coronal Fe, Mg and Si, and a possible enhancement of the
elements with lowest FIP (Ca and Ni), although still consistent with
the photospheric values (Fig.~\ref{figabund2}). The determination of
the photospheric abundance of elements with larger FIP would be
desirable in order to confirm any trend.

\section{Discussion and conclusions}\label{sec:conclusions}

The results found in the present sample show that the issue of coronal
abundances in stars other than the Sun is complex and still poorly
understood. Solar coronal abundance is by itself a complex issue, with
the actual values depending on which part of the corona is
observed (with coronal holes, active regions and flares showing quite
different abundance patterns). Results obtained from full disk spectra
(thus comparable to stellar spectra) show enhancements by at least a
factor of 3 in elements with ${\rm FIP}\la 10$~eV \citep{lam95}.
Young solar active regions show coronal abundance compatible with the
photospheric values, but they soon develop a FIP bias that reaches
average coronal values in two to three days \citep{she96,wid01}. The
FIP effect that is observed in the solar corona seems to disappear with
increasing activity level of the star, as it has been shown in cases
like II~Peg \citep{huen01}, AR~Lac \citep{huen03}, AB~Dor
\citep{sanz03}, $\lambda$~And and V851~Cen.  A solar-like FIP effect
appears to be present in $\alpha$~Cen~A \citep{raa03}, a star very
similar to the Sun in many physical parameters, as well as in other
solar analogs with low levels of activity, such as $\pi^1$~UMa and
$\chi^1$~Ori \citep{gud02}, although only the Fe photospheric
abundance is known for these two objects (so that abundance patterns
assume a solar-like mix). A mid-activity level star, $\epsilon$~Eri,
could be showing the progressive disappearance of the FIP effect, with
elements like Mg and Fe already showing photospheric abundances in the
corona, and leaving a marginal coronal abundance enhancement only for
Ca and Ni.  Finally, Procyon seems to be quite different from the Sun,
with no FIP-related trends evident, although there could be an
increment of Fe abundance with the coronal temperature relative to
other elements with higher FIP (see Sect.~\ref{sec:procyon}). The FIP
effect detected in the solar corona above quiet regions grows with
temperature, in a range similar to that covered here for Procyon
\citep{feld00}.

The interpretation of the metal depletion in the corona of active
stars that has been given in the past from initial results is somewhat
in contradiction with the present results.  Many of the stars for
which a ``metal abundance deficiency syndrome'' (MAD) has been claimed
do not have measurements of the photospheric abundance available (or
they are limited to Fe abundance determinations), and published work
is limited to the comparison of stellar coronal patterns with solar
photospheric abundances (a rather arbitrary practice). The case of
$\lambda$~And is quite remarkable, with compatible coronal and
photospheric abundances for all elements tested, and a definitely
non-solar photospheric abundance pattern. V851~Cen, with similar
coronal thermal structure (Fig.~\ref{emds}) shows a marginal coronal
abundance depletion. Other cases like II Peg \citep{huen01} or AR Lac
\citep{huen03} show also moderate depletion when they are compared
with the respective photospheric values, but strong metal depletion
seems to be clear only for AB~Dor \citep[][and references
therein]{sanz03}.  However AB~Dor is a fast rotator, and the only
available determination of photospheric abundances, carried out by
\citet{vil87}, needs further confirmation given the difficulty for the
measurements in a spectrum so much affected by rotational broadening.

While the absence of FIP effect in active stars, at least at the solar
level, is well supported at this point, the presence of a general
``MAD effect'' or an ``inverse FIP effect'' increasing with activity
is not yet clearly established, given the discrepancies found between
stars with quite similar levels of activity and thermal coronal
structure like AB~Dor, $\lambda$~And and V851 Cen.  Moreover, the
abundance pattern could also be changing in different parts of the
corona, depending on the coronal temperature, as present in Procyon.
A larger sample of stars with both photospheric and coronal
abundances, determined in a homogeneous way, would be necessary in
order to establish a general trend between coronal abundances and
activity levels.


\begin{acknowledgements}
  We acknowledge support by the Marie Curie Fellowships Contract No.
  HPMD-CT-2000-00013.  We have made use of data obtained through the
  \emph{Chandra} data archive, operated by the Smithsonian
  Astrophysical Observatory for NASA. This research has also made use
  of NASA's Astrophysics Data System Abstract Service. We are grateful
  to the referee, Dr. J. M. Laming, for the careful reading and useful
  comments brought to the manuscript.
\end{acknowledgements}

\Online
\begin{table*}
\caption{Chandra/LETG line fluxes of Procyon$^a$}\label{tab:flprocyon}
\tabcolsep 3.pt
\begin{scriptsize}
\begin{tabular}{lrcrrrl}
\hline \hline
 Ion & {$\lambda$$_{\mathrm {model}}$} &  
 log $T_{\mathrm {max}}$ & $F_{\mathrm {obs}}$ & S/N & ratio & Blends \\ 
\hline
\ion{Fe}{xvii} &  12.1240 & 6.8 & 6.03e-15 &   3.6 & -0.18 & \ion{Ne}{x} 12.1321, \ion{Ne}{x} 12.1375 \\
\ion{Ne}{ix} &  13.4473 & 6.6 & 2.79e-14 &   8.3 & -0.34 &  \\
\ion{Ne}{ix} &  13.6990 & 6.6 & 1.53e-14 &   6.2 & -0.36 & \ion{Ne}{ix} 13.7130 \\
\ion{Fe}{xvii} &  15.0140 & 6.7 & 2.71e-14 &   8.8 & -0.15 &  \\
\ion{Fe}{xvii} &  15.2610 & 6.7 & 2.45e-14 &   8.5 &  0.00 & \ion{O }{viii} 15.1760, 15.1765, \ion{Fe}{xvii} 15.2509, 15.2615, 15.2797 \\
\ion{O }{viii} &  16.0055 & 6.5 & 3.04e-14 &   8.3 &  0.07 & \ion{O }{viii} 16.0067 \\
\ion{Fe}{xvii} &  16.7800 & 6.7 & 1.44e-14 &   6.8 & -0.12 &  \\
\ion{Fe}{xvii} &  17.0510 & 6.7 & 5.22e-14 &  12.0 &  0.14 & \ion{Fe}{xvii} 17.0960 \\
\ion{O }{vii} &  17.3960 & 6.4 & 5.25e-15 &   3.8 & -0.01 &  \\
\ion{O }{vii} &  17.7680 & 6.4 & 1.69e-14 &   6.9 &  0.13 &  \\
\ion{O }{vii} &  18.6270 & 6.3 & 4.76e-14 &  12.3 &  0.04 &  \\
\ion{O }{viii} &  18.9671 & 6.5 & 1.88e-13 &  24.7 & -0.05 & \ion{O }{viii} 18.9725 \\
\ion{N }{vii} &  20.9095 & 6.3 & 1.26e-14 &   5.4 &  0.15 & \ion{N }{vii} 20.9106 \\
\ion{O }{vii} &  21.6015 & 6.3 & 2.66e-13 &  25.1 & -0.14 &  \\
\ion{O }{vii} &  21.8036 & 6.3 & 7.20e-14 &  13.0 &  0.02 &  \\
\ion{O }{vii} &  22.0977 & 6.3 & 2.42e-13 &  24.0 &  0.03 &  \\
\ion{N }{vii} &  24.7792 & 6.3 & 7.00e-14 &  13.6 & -0.02 & \ion{N }{vii} 24.7846 \\
\ion{N }{vi} &  24.8980 & 6.2 & 5.64e-15 &   3.9 & -0.12 &  \\
\ion{C }{vi} &  28.4652 & 6.2 & 2.85e-14 &   9.0 &  0.02 &  \\
\ion{N }{vi} &  28.7870 & 6.2 & 5.09e-14 &  11.9 & -0.03 &  \\
\ion{N }{vi} &  29.0843 & 6.1 & 1.97e-14 &   7.4 &  0.17 &  \\
\ion{N }{vi} &  29.5347 & 6.1 & 1.69e-14 &   6.7 & -0.23 &  \\
\ion{Ca}{xi} &  30.4710 & 6.3 & 2.19e-14 &   5.1 &  0.01 & \ion{S }{xiv} 30.4270 \\
No id. &  31.0340 & \ldots  & 1.69e-14 &   5.0 & \ldots & \ion{Si}{xii} 31.015 \\
\ion{C }{vi} &  33.7342 & 6.1 & 2.44e-13 &  25.8 & -0.10 & \ion{C }{vi} 33.7396 \\
\ion{C }{v} &  34.9728 & 6.0 & 1.85e-14 &   7.2 &  0.17 &  \\
\ion{S }{xiii} &  35.6670 & 6.4 & 2.66e-14 &   8.6 &  0.09 & \ion{Ca}{xi} 35.6340, 35.7370 \\
\ion{S }{xii} &  36.3980 & 6.4 & 1.55e-14 &   6.5 &  0.06 &  \\
\ion{S }{xii} &  36.5730 & 6.4 & 1.01e-14 &   5.2 & -0.17 & \ion{S }{xii} 36.5640 \\
\ion{S }{xiii} &  37.5980 & 6.4 & 1.24e-14 &   5.5 &  0.20 & \ion{S }{xii} 37.5670 \\
\ion{S }{xi} &  39.2400 & 6.3 & 4.48e-14 &  10.2 &  0.06 & \ion{S }{xi} 39.3000,  39.3200 \\
\ion{C }{v} &  40.2674 & 6.0 & 1.06e-13 &  12.2 &  0.13 &  \\
No id. &  40.7250 & \ldots  & 9.27e-14 &   9.2 & \ldots &  \ion{C}{v} 40.731 \\
\ion{C }{v} &  41.4715 & 6.0 & 3.68e-14 &   5.7 & -0.26 & \ion{Ar}{ix} 41.4760 \\
No id. &  42.5330 & \ldots  & 5.48e-14 &   7.7 & \ldots & \ion{S}{x} 42.543 \\
\ion{Si}{xi} &  43.7630 & 6.2 & 3.69e-14 &  10.8 & -0.02 &  \\
\ion{Si}{xii} &  44.0190 & 6.3 & 1.38e-14 &   6.8 & -0.24 &  \\
\ion{Si}{xii} &  44.1650 & 6.3 & 4.44e-14 &  12.2 & -0.18 & \ion{Mg}{x} 44.0500, \ion{Si}{xi} 44.0690, \ion{Si}{xii} 44.1780 \\
No id. &  44.4110 & \ldots  & 6.59e-15 &   4.7 & \ldots &  \\
\ion{Si}{xii} &  45.6910 & 6.3 & 1.09e-14 &   9.1 & -0.34 &  \\
\ion{Si}{xi} &  46.2980 & 6.2 & 1.47e-14 &  10.5 &  0.30 &  \\
\ion{Si}{xi} &  46.3990 & 6.2 & 1.61e-14 &  11.0 & -0.15 & \ion{Si}{xi} 46.4090 \\
\ion{Fe}{xvi} &  46.7180 & 6.5 & 9.39e-15 &   8.4 &  0.25 & \ion{Fe}{xvi} 46.6610 \\
\ion{S }{xi} &  47.0300 & 6.3 & 8.36e-15 &   5.9 & -0.08 & \ion{Si}{xi} 46.9960, 47.0770, \ion{Si}{x} 47.0450 \\
No id. &  47.2500 & \ldots  & 1.66e-14 &   8.3 & \ldots & \ion{S}{ix} 47.249 \\
No id. &  47.4000 & \ldots  & 1.23e-14 &   7.1 & \ldots & \ion{S}{ix} 47.433 \\
\ion{Si}{x} &  47.4890 & 6.2 & 1.29e-14 &   7.3 &  0.31 &  \\
No id. &  47.6240 &\ldots   & 1.77e-14 &   8.5 & \ldots & \ion{S}{x} 47.655 \\
No id. &  47.7500 & \ldots  & 1.12e-14 &   6.8 & \ldots & \ion{S}{x} 47.791 \\
No id. &  47.8790 &\ldots   & 1.47e-14 &   7.8 & \ldots & \ion{Si}{xi} 47.899 \\
\ion{Ar}{ix} &  48.7390 & 6.1 & 7.53e-15 &   7.5 &  0.00 & \ion{Si}{x} 48.6760, 48.7890 \\
\ion{Si}{xi} &  49.2220 & 6.2 & 6.12e-14 &  21.7 & -0.11 &  \\
No id. &  49.3340 & \ldots  & 1.48e-14 &  10.7 & \ldots & \ion{S}{ix} 49.328 \\
\ion{Si}{x} &  49.7010 & 6.2 & 7.39e-15 &   7.6 &  0.14 & \ion{Si}{x} 49.7510 \\
\ion{Fe}{xvi} &  50.3500 & 6.5 & 1.65e-14 &  11.4 &  0.20 & \ion{Si}{x} 50.3160, 50.3330, 50.3590 \\
\ion{Si}{x} &  50.5240 & 6.2 & 6.60e-14 &  22.8 &  0.12 & \ion{Fe}{xvi} 50.5550 \\
\ion{Si}{x} &  50.7030 & 6.2 & 4.43e-14 &  18.7 & -0.24 & \ion{Si}{x} 50.6910 \\
\ion{Si}{xi} &  52.2980 & 6.2 & 2.53e-14 &  13.8 & -0.42 & \ion{Si}{xi} 52.2690, \ion{Al}{xi} 52.2990 \\
\ion{Al}{xi} &  52.4460 & 6.2 & 9.05e-15 &   8.3 & -0.00 &  \\
\ion{Si}{x} &  52.6110 & 6.2 & 9.34e-15 &   8.5 & -0.08 & \ion{Si}{x} 52.6120 \\
No id. &  52.7600 & \ldots  & 1.55e-14 &  10.8 & \ldots & \ion{S}{viii} 52.756 \\
\ion{Fe}{xv} &  52.9110 & 6.4 & 9.92e-15 &   8.8 & -0.04 &  \\
No id. &  53.0480 & \ldots  & 7.36e-15 &   7.5 & \ldots &  \\
\ion{Fe}{xv} &  53.1100 & 6.4 & 6.60e-15 &   7.0 &  0.22 & \ion{Si}{x} 53.1530 \\
\ion{Si}{x} &  54.5990 & 6.2 & 1.10e-14 &   8.9 &  0.32 &  \\
\ion{Si}{ix} &  55.1160 & 6.1 & 3.04e-14 &  14.7 &  0.40 & \ion{Si}{ix} 55.0390, 55.0940, \ion{Ni}{xvii} 55.0390 \\
\ion{Si}{ix} &  55.3050 & 6.1 & 8.40e-14 &  24.2 &  0.05 & \ion{Si}{ix} 55.2340, 55.3830, 55.4010 \\
No id. &  56.9000 & \ldots  & 1.53e-14 &   7.1 & \ldots & \ion{Si}{x} 56.08 \\
\ion{Fe}{xv} &  59.4050 & 6.3 & 8.68e-15 &   5.2 & -0.52 &  \\
\ion{S }{vii} &  60.1610 & 5.8 & 1.13e-14 &   5.9 &  0.35 & \ion{Si}{x} 60.1180 \\
No id. &  61.0500 & \ldots  & 5.52e-14 &  13.1 & \ldots & \ion{Si}{viii} 61.019 \\
\ion{Si}{ix} &  61.6490 & 6.1 & 3.90e-14 &  11.0 &  0.49 & \ion{Si}{ix} 61.6000, 61.6970 \\
No id. &  61.8500 &\ldots   & 2.78e-14 &   9.3 & \ldots & \ion{Si}{ix} 61.852 \\
No id. &  61.9300 &\ldots   & 2.81e-14 &   9.3 & \ldots & \ion{Si}{viii} 61.914 \\
\ion{Si}{x} &  62.6950 & 6.2 & 2.17e-14 &   8.2 & -0.18 & \ion{Mg}{ix} 62.6610, 62.7510 \\
\ion{Fe}{xvi} &  62.8790 & 6.5 & 1.92e-14 &   7.7 &  0.04 &  \\
\ion{Si}{ix} &  62.9750 & 6.1 & 9.72e-15 &   5.5 &  0.61 &  \\
\ion{Mg}{x} &  63.1520 & 6.1 & 2.08e-14 &   8.0 & -0.19 &  \\
\ion{Mg}{x} &  63.2950 & 6.1 & 3.82e-14 &  10.9 & -0.23 & \ion{Mg}{x} 63.3110 \\
\ion{Fe}{xvi} &  63.7190 & 6.5 & 2.25e-14 &   8.4 & -0.21 &  \\
\ion{Fe}{xv} &  63.9580 & 6.3 & 1.94e-14 &   7.8 &  0.26 &  \\
No id. &  64.1230 & \ldots  & 2.52e-14 &   8.8 & \ldots &  \\
\ion{Mg}{x} &  65.6730 & 6.1 & 9.77e-15 &   7.8 & -0.22 &  \\
\ion{Mg}{x} &  65.8450 & 6.1 & 2.69e-14 &  13.3 & -0.11 & \ion{Ne}{viii} 65.8940 \\
\hline
\end{tabular}
\end{scriptsize}
\end{table*}
\setcounter{table}{3}
\begin{table*}
\caption{(cont). Chandra/LETG line fluxes of Procyon$^a$}
\tabcolsep 3.pt
\begin{scriptsize}
\begin{tabular}{lrcrrrl}
\hline \hline
 Ion & {$\lambda$$_{\mathrm {model}}$} & 
 log $T_{\mathrm {max}}$ & $F_{\mathrm {obs}}$ & S/N & ratio & Blends \\ 
\hline
\ion{Fe}{xvi} &  66.2630 & 6.5 & 9.90e-15 &   7.9 & -0.07 &  \\
No id. &  66.3130 & \ldots  & 9.61e-15 &   8.0 & \ldots & \ion{Ne}{viii} 66.330 \\
\ion{Fe}{xvi} &  66.3680 & 6.5 & 1.58e-14 &  10.0 & -0.04 &  \\
\ion{Mg}{ix} &  67.1350 & 6.0 & 1.37e-14 &   6.3 &  0.30 & \ion{Mg}{ix} 67.1410 \\
\ion{Mg}{ix} &  67.2390 & 6.0 & 2.55e-14 &   8.6 &  0.19 & \ion{Mg}{ix} 67.2460 \\
\ion{Ne}{viii} &  67.3820 & 5.9 & 1.84e-14 &   7.3 &  0.33 & \ion{Si}{x} 67.3550, \ion{Ne}{viii} 67.3860 \\
\ion{Fe}{xv} &  69.6490 & 6.3 & 5.09e-14 &  18.0 & -0.18 &  \\
No id. &  69.8730 & \ldots  & 3.50e-14 &  14.8 & \ldots & \ion{Si}{viii} 69.825 \\
\ion{Fe}{xv} &  70.0540 & 6.3 & 2.90e-14 &  13.5 &  0.34 & \ion{Fe}{xv} 69.9450, 69.9870 \\
No id. &  71.9110 & \ldots  & 1.99e-14 &  11.1 & \ldots & \ion{Mg}{ix} 71.901 \\
\ion{S }{vii} &  72.0280 & 5.8 & 1.82e-14 &  10.6 &  0.60 & \ion{Mg}{ix} 72.0280 \\
\ion{Mg}{ix} &  72.3120 & 6.0 & 3.48e-14 &  14.7 & -0.12 &  \\
\ion{S }{vii} &  72.6630 & 5.8 & 1.50e-14 &   9.7 &  1.01 &  \\
\ion{S }{vii} &  72.8980 & 5.8 & 1.47e-14 &   9.5 &  0.87 &  \\
\ion{Fe}{xv} &  73.4730 & 6.3 & 1.08e-14 &   8.2 & -0.43 & \ion{Ne}{viii} 73.4750 \\
\ion{Ne}{viii} &  73.5630 & 5.9 & 1.07e-14 &   8.1 &  0.19 & \ion{Ne}{viii} 73.5650 \\
\ion{Mg}{viii} &  74.8580 & 5.9 & 2.39e-14 &  11.9 &  0.10 &  \\
\ion{Mg}{viii} &  75.0440 & 5.9 & 3.01e-14 &  13.4 & -0.02 & \ion{Mg}{viii} 75.0340 \\
No id. &  76.0310 & \ldots  & 1.96e-14 &  10.7 & \ldots & \ion{Fe}{x} 76.006 \\
\ion{Mg}{ix} &  77.7370 & 6.0 & 2.86e-14 &  12.8 & -0.12 &  \\
No id. &  77.8340 & \ldots  & 1.16e-14 &   8.1 & \ldots &  \\
No id. &  79.4950 & \ldots  & 1.23e-14 &   8.4 & \ldots & \ion{Fe}{xii} 79.488 \\
No id. &  80.5320 & \ldots  & 1.38e-14 &   8.7 & \ldots & \ion{Fe}{xii} 80.510 \\
No id. &  82.7000 & \ldots  & 2.92e-14 &   8.6 & \ldots & \ion{Fe}{xii} 82.744 \\
No id. &  83.5980 & \ldots  & 1.02e-14 &   7.3 & \ldots & \ion{Mg}{viii} 83.587 \\
\ion{Mg}{vii} &  83.9100 & 5.9 & 4.07e-15 &   4.6 & -0.01 &  \\
No id. &  83.9960 & \ldots  & 1.82e-14 &   9.7 & \ldots & \ion{Mg}{vii} 84.025 \\
No id. &  86.8220 & \ldots  & 1.94e-14 &  10.1 & \ldots & \ion{Mg}{viii} 86.847 \\
\ion{Mg}{viii} &  87.0210 & 5.9 & 1.50e-14 &   8.8 &  0.47 &  \\
\ion{Ne}{viii} &  88.0820 & 5.8 & 5.06e-14 &  16.2 &  0.06 & \ion{Ne}{viii} 88.1190 \\
No id. &  91.8360 & \ldots  & 1.97e-14 &  10.2 & \ldots & \ion{Ni}{x} 91.790 \\
No id. &  94.0510 & \ldots  & 3.76e-14 &  13.9 & \ldots & \ion{Fe}{x} 94.012\\
No id. &  96.0420 & \ldots  & 3.18e-14 &  12.5 & \ldots & \ion{Si}{vi} 96.022 \\
No id. &  96.1430 & \ldots  & 3.22e-14 &  12.5 & \ldots & \ion{Fe}{x}  96.122 \\
\ion{Ne}{vii} &  97.4950 & 5.8 & 1.37e-14 &   8.0 &  0.27 &  \\
\ion{Ne}{viii} &  98.1150 & 5.8 & 2.78e-14 &  11.4 &  0.06 &  \\
\ion{Ne}{viii} &  98.2600 & 5.8 & 5.25e-14 &  15.6 &  0.08 &  \\
No id. & 100.6000 & \ldots  & 2.42e-14 &  10.5 & \ldots & \ion{Mg}{viii} 100.597 \\
No id. & 102.1000 & \ldots  & 6.31e-15 &   5.3 & \ldots &  \\
\ion{Ne}{viii} & 102.9110 & 5.8 & 1.41e-14 &   8.0 &  0.06 &  \\
\ion{Ne}{viii} & 103.0850 & 5.8 & 2.40e-14 &  10.4 & -0.01 &  \\
No id. & 103.5740 & \ldots  & 3.38e-14 &  12.3 & \ldots & \ion{Fe}{ix} 103.566 \\
No id. & 103.9370 & \ldots  & 1.42e-14 &   8.0 & \ldots &  \\
No id. & 105.2350 & \ldots  & 2.57e-14 &  10.8 & \ldots & \ion{Fe}{ix} 105.208 \\
No id. & 111.2550 & \ldots  & 1.93e-14 &   9.4 & \ldots & \ion{Ca}{x} 111.198 \\
No id. & 113.7970 & \ldots  & 1.52e-14 &   8.3 & \ldots & \ion{Fe}{viii} 113.763 \\
\ion{O }{vi} & 115.8210 & 5.5 & 1.34e-14 &   7.9 &  0.32 & \ion{O }{vi} 115.8300 \\
\ion{Ne}{vii} & 116.6930 & 5.8 & 1.13e-14 &   7.2 & -0.00 &  \\
No id. & 127.7000 & \ldots  & 1.45e-14 &   6.5 & \ldots & \ion{Ne}{vii} 127.663 \\
\ion{Fe}{viii} & 130.9410 & 5.7 & 1.23e-14 &   6.0 & -0.03 &  \\
No id. & 131.2800 & \ldots  & 2.32e-14 &   8.2 & \ldots & \ion{Fe}{viii} 131.24 \\
No id. & 148.4290 & \ldots  & 1.36e-13 &  20.3 & \ldots & \ion{Ni}{xi} 148.402 \\
\ion{O }{vi} & 150.0890 & 5.5 & 7.13e-14 &  14.7 &  0.43 & \ion{O }{vi} 150.1250 \\
\ion{Ni}{xii} & 152.1540 & 6.2 & 7.13e-14 &  14.7 &  0.01 &  \\
\ion{Ni}{xii} & 154.1620 & 6.2 & 3.60e-14 &  10.4 &  0.02 &  \\
\ion{Ni}{xiii} & 157.7290 & 6.2 & 3.11e-14 &   9.5 & -0.10 & \ion{Ni}{xii} 157.8130 \\
No id. & 158.4290 & \ldots  & 2.20e-14 &   8.0 & \ldots & \ion{Ni}{x} 158.377 \\
No id. & 160.0590 & \ldots  & 2.26e-14 &   7.9 & \ldots & \ion{Ni}{x} 159.977 \\
\ion{Ni}{xii} & 160.5550 & 6.2 & 1.75e-14 &   6.9 &  0.06 &  \\
No id. & 167.5560 & \ldots  & 4.32e-14 &   9.7 & \ldots & \ion{Fe}{viii} 167.486 \\
No id. & 168.2090 & \ldots  & 6.75e-14 &  11.8 & \ldots & \ion{Fe}{viii} 168.172 \\
\ion{Fe}{ix} & 171.0730 & 5.9 & 8.75e-13 &  29.7 & -0.17 &  \\
\hline
\end{tabular}

{$^a$ Line fluxes (in erg cm$^{-2}$ s$^{-1}$) 
  measured in Chandra/LETG Procyon summed spectra. 
  log $T_{\mathrm {max}}$ indicates the maximum
  temperature (K) of formation of the line (unweighted by the
  EMD). ``Ratio'' is the log~($F_{\mathrm {obs}}$/$F_{\mathrm {pred}}$) 
  of the line. 
  Blends amounting to more than 5\% of the total flux for each line are
  indicated. For some lines not identified in APED, a tentative identification is suggested in the ``Blends'' column based on \citet{kelly}.}
\end{scriptsize}
\end{table*}

\begin{table*}
\caption{Chandra/LETG line fluxes of $\epsilon$~Eri$^a$}\label{tab:fleeri}
\tabcolsep 3.pt
\begin{scriptsize}
\begin{tabular}{lrcrrrl}
\hline \hline
 Ion & {$\lambda$$_{\mathrm {model}}$} &  
 log $T_{\mathrm {max}}$ & $F_{\mathrm {obs}}$ & S/N & ratio & Blends \\ 
\hline
\ion{Si}{xiii} &   6.6479 & 7.0 & 4.57e-14 &   8.0 & -0.07 & \ion{Si}{xiii}  6.6882 \\
\ion{Si}{xiii} &   6.7403 & 7.0 & 2.35e-14 &   5.7 & -0.02 & \ion{Si}{xiii}  6.7388,  6.7432 \\
\ion{Mg}{xii} &   8.4192 & 7.0 & 3.00e-14 &   6.8 & -0.10 & \ion{Mg}{xii}  8.4246 \\
\ion{Mg}{xi} &   9.1687 & 6.8 & 4.95e-14 &   8.4 & -0.25 &  \\
\ion{Mg}{xi} &   9.2312 & 6.8 & 7.83e-14 &  10.5 &  0.12 & \ion{Mg}{xi}  9.3143 \\
\ion{Ne}{x} &   9.7080 & 6.8 & 1.65e-14 &   4.8 & -0.00 & \ion{Ne}{x}  9.7080,  9.7085, \ion{Fe}{xix}  9.6938 \\
\ion{Ne}{x} &  10.2385 & 6.8 & 3.31e-14 &   6.8 & -0.17 & \ion{Ne}{x} 10.2396 \\
\ion{Fe}{xvii} &  11.2540 & 6.8 & 4.85e-14 &   8.6 & -0.01 & \ion{Fe}{xviii} 11.3260 \\
\ion{Fe}{xviii} &  11.5270 & 6.9 & 4.99e-14 &   8.9 & -0.05 & \ion{Fe}{xviii} 11.5270, \ion{Ne}{ix} 11.5440 \\
\ion{Ne}{x} &  12.1320 & 6.8 & 3.15e-13 &  22.8 &  0.14 & \ion{Ne}{x} 12.1321 \\
\ion{Fe}{xxi} &  12.2840 & 7.0 & 7.90e-14 &  11.5 & -0.08 & \ion{Fe}{xvii} 12.2660 \\
\ion{Ni}{xix} &  12.4350 & 6.9 & 6.62e-14 &  10.6 &  0.06 & \ion{Fe}{xxi} 12.3930 \\
\ion{Fe}{xx} &  12.5760 & 7.0 & 1.38e-14 &   4.9 &  0.26 & \ion{Fe}{xx} 12.5760, \ion{Fe}{xxi} 12.5698 \\
\ion{Fe}{xxi} &  12.6490 & 7.0 & 1.20e-14 &   4.5 & -0.04 & \ion{Ni}{xix} 12.6560 \\
No id. &  12.7550 & \ldots & 1.30e-14 &   4.8 & \ldots & \ion{Fe}{xxii} 12.7540 \\
\ion{Fe}{xx} &  12.8240 & 7.0 & 4.33e-14 &   8.7 & -0.17 & \ion{Fe}{xx} 12.8460, 12.8640 \\
\ion{Fe}{xx} &  12.9650 & 7.0 & 4.09e-14 &   8.5 &  0.24 & \ion{Fe}{xx} 12.9120, 12.9920, \ion{Fe}{xix} 12.9330 \\
\ion{Ne}{ix} &  13.4473 & 6.6 & 4.98e-13 &  30.6 &  0.07 & \ion{Fe}{xix} 13.5180, \ion{Ne}{ix} 13.5531 \\
\ion{Ne}{ix} &  13.6990 & 6.6 & 2.01e-13 &  19.6 &  0.13 &  \\
\ion{Fe}{xvii} &  13.8250 & 6.8 & 1.18e-13 &  15.1 &  0.03 & \ion{Fe}{xix} 13.7458, 13.7950, \ion{Ni}{xix} 13.7790, \ion{Fe}{xvii} 13.8920 \\
\ion{Fe}{xviii} &  13.9530 & 6.9 & 3.45e-14 &   8.2 &  0.40 & \ion{Fe}{xix} 13.9549, 13.9551, \ion{Fe}{xx} 13.9620 \\
\ion{Ni}{xix} &  14.0430 & 6.8 & 4.48e-14 &   9.4 & -0.06 & \ion{Fe}{xxi} 14.0080, \ion{Ni}{xix} 14.0770 \\
\ion{Fe}{xviii} &  14.2080 & 6.9 & 1.09e-13 &  14.7 & -0.17 & \ion{Fe}{xviii} 14.2080 \\
\ion{Fe}{xviii} &  14.2560 & 6.9 & 4.82e-14 &   9.8 &  0.09 & \ion{Fe}{xviii} 14.2560, \ion{Fe}{xx} 14.2670 \\
\ion{Fe}{xviii} &  14.3430 & 6.9 & 9.73e-14 &  14.0 &  0.06 & \ion{Fe}{xviii} 14.3730, 14.4250, 14.4392 \\
\ion{Fe}{xviii} &  14.5710 & 6.9 & 6.44e-14 &  11.6 &  0.09 & \ion{Fe}{xviii} 14.5340, 14.6011 \\
\ion{Fe}{xix} &  14.6640 & 6.9 & 2.65e-14 &   7.5 &  0.26 & \ion{Fe}{xviii} 14.6884 \\
\ion{Fe}{xvii} &  15.0140 & 6.7 & 6.12e-13 &  36.4 & -0.05 &  \\
\ion{Fe}{xvii} &  15.2610 & 6.7 & 3.56e-13 &  28.1 &  0.13 & \ion{O }{viii} 15.1760 \\
\ion{Fe}{xvii} &  15.4530 & 6.7 & 6.57e-14 &  12.0 &  0.36 & \ion{Fe}{xviii} 15.4940 \\
\ion{Fe}{xviii} &  15.6250 & 6.8 & 3.52e-14 &   8.9 & -0.05 &  \\
\ion{Fe}{xviii} &  15.7590 & 6.8 & 1.35e-14 &   5.6 &  0.38 &  \\
\ion{Fe}{xviii} &  15.8240 & 6.8 & 3.61e-14 &   9.1 & -0.03 & \ion{Fe}{xviii} 15.8700 \\
\ion{Fe}{xviii} &  15.9310 & 6.8 & 1.60e-14 &   6.0 &  0.24 &  \\
\ion{O }{viii} &  16.0055 & 6.5 & 1.75e-13 &  20.0 &  0.03 & \ion{Fe}{xviii} 16.0040, \ion{O }{viii} 16.0067 \\
\ion{Fe}{xviii} &  16.0710 & 6.8 & 9.38e-14 &  14.8 & -0.00 & \ion{Fe}{xix} 16.1100, \ion{Fe}{xviii} 16.1590 \\
\ion{Fe}{xvii} &  16.3500 & 6.7 & 1.93e-14 &   6.9 &  0.09 & \ion{Fe}{xviii} 16.3200, \ion{Fe}{xix} 16.3414 \\
\ion{Fe}{xvii} &  16.7800 & 6.7 & 3.12e-13 &  27.5 &  0.01 &  \\
\ion{Fe}{xvii} &  17.0510 & 6.7 & 8.11e-13 &  41.0 &  0.13 & \ion{Fe}{xvii} 17.0960 \\
\ion{O }{vii} &  18.6270 & 6.3 & 4.60e-14 &  10.5 & -0.07 &  \\
\ion{Ca}{xviii} &  18.7320 & 6.9 & 6.14e-15 &   3.8 & -0.03 & \ion{Ca}{xviii} 18.6910, \ion{Ar}{xvi} 18.6960 \\
No id. &  18.7870 & \ldots & 9.36e-15 &   4.8 & \ldots &  \\
\ion{O }{viii} &  18.9671 & 6.5 & 7.31e-13 &  42.4 & -0.02 & \ion{O }{viii} 18.9725 \\
\ion{N }{vii} &  20.9095 & 6.3 & 1.04e-14 &   4.2 & -0.08 & \ion{N }{vii} 20.9106 \\
\ion{Ca}{xvii} &  21.1980 & 6.8 & 2.31e-14 &   6.3 &  0.27 &  \\
\ion{Ca}{xvi} &  21.4500 & 6.7 & 9.30e-15 &   4.1 &  0.00 &  \\
\ion{O }{vii} &  21.6015 & 6.3 & 3.98e-13 &  26.6 & -0.02 &  \\
No id. &  21.7100 & \ldots & 2.18e-14 &   6.2 & \ldots &  \\
\ion{O }{vii} &  21.8036 & 6.3 & 8.64e-14 &  12.4 &  0.04 &  \\
No id. &  22.0220 & \ldots & 1.61e-14 &   5.4 & \ldots &  \\
\ion{O }{vii} &  22.0977 & 6.3 & 2.64e-13 &  21.8 &  0.05 &  \\
\ion{S }{xiv} &  24.2850 & 6.5 & 1.89e-14 &   6.1 &  0.58 & \ion{S }{xiv} 24.2890 \\
\ion{N }{vii} &  24.7792 & 6.3 & 8.76e-14 &  13.2 & -0.02 & \ion{N }{vii} 24.7846 \\
\ion{N }{vi} &  24.8980 & 6.2 & 1.14e-14 &   4.8 &  0.04 & \ion{Ar}{xvi} 24.8540 \\
\ion{Ar}{xvi} &  24.9910 & 6.7 & 1.03e-14 &   4.5 & -0.00 &  \\
\ion{C }{vi} &  28.4652 & 6.2 & 2.41e-14 &   7.2 & -0.05 & \ion{C }{vi} 28.4663 \\
\ion{N }{vi} &  28.7870 & 6.2 & 2.24e-14 &   6.9 & -0.14 &  \\
\ion{N }{vi} &  29.0843 & 6.1 & 2.14e-14 &   6.7 &  0.41 &  \\
\ion{N }{vi} &  29.5347 & 6.1 & 1.45e-14 &   5.4 &  0.03 &  \\
\ion{S }{xiv} &  30.4270 & 6.5 & 3.95e-14 &   8.3 & -0.28 & \ion{S }{xiv} 30.4690, \ion{Ca}{xi} 30.4710 \\
No id. &  32.2410 & \ldots & 2.76e-14 &   7.6 & \ldots &  \\
\ion{S }{xiv} &  32.4160 & 6.5 & 2.14e-14 &   6.7 &  0.05 &  \\
\ion{S }{xiv} &  32.5600 & 6.5 & 3.20e-14 &   8.1 & -0.08 & \ion{S }{xiv} 32.5750 \\
\ion{S }{xiv} &  33.5490 & 6.5 & 2.45e-14 &   7.1 &  0.13 &  \\
\ion{C }{vi} &  33.7342 & 6.1 & 1.76e-13 &  19.1 & -0.01 & \ion{C }{vi} 33.7396 \\
\ion{S }{xiii} &  35.6670 & 6.4 & 3.66e-14 &   8.8 & -0.17 & \ion{Ca}{xi} 35.6340, 35.7370 \\
\ion{S }{xii} &  36.3980 & 6.4 & 3.44e-14 &   8.4 &  0.19 &  \\
\ion{S }{xiii} &  37.5980 & 6.4 & 1.81e-14 &   5.8 & -0.09 &  \\
No id. &  37.9500 & \ldots & 3.67e-14 &   8.2 & \ldots & (\ion{O}{viii} 18.97, 2$^{\rm nd}$ order) \\
\ion{C }{v} &  40.2674 & 6.0 & 4.11e-14 &   6.7 &  0.29 & \ion{S }{xi} 40.2730 \\
\ion{S }{xii} &  43.6510 & 6.4 & 9.55e-15 &   6.7 &  0.45 &  \\
\ion{Si}{xi} &  43.7630 & 6.2 & 2.48e-14 &  11.7 &  0.22 &  \\
\ion{Si}{xii} &  44.0190 & 6.3 & 3.54e-14 &  14.0 &  0.06 & \ion{Mg}{x} 44.0500, \ion{Si}{xi} 44.0690 \\
\ion{Si}{xii} &  44.1650 & 6.3 & 4.95e-14 &  16.6 &  0.06 & \ion{Si}{xii} 44.1780 \\
\ion{Si}{xii} &  45.5210 & 6.3 & 1.05e-14 &   7.7 &  0.01 &  \\
\ion{Si}{xii} &  45.6910 & 6.3 & 1.98e-14 &  10.6 & -0.02 &  \\
\ion{Ar}{ix} &  48.7390 & 6.1 & 6.69e-15 &   6.2 & -0.01 & \ion{Si}{x} 48.6760, 48.7890 \\
\ion{Si}{xi} &  49.2220 & 6.2 & 3.08e-14 &  13.4 & -0.09 & \ion{Ar}{ix} 49.1860, \ion{Fe}{xvii} 49.2623 \\
\ion{Fe}{xvi} &  50.3500 & 6.5 & 3.51e-14 &  14.2 &  0.27 & \ion{Fe}{xvii} 50.3419 \\
\ion{Si}{x} &  50.5240 & 6.2 & 2.88e-14 &  12.7 &  0.10 & \ion{Fe}{xvi} 50.5550 \\
\ion{Si}{x} &  50.7030 & 6.2 & 1.83e-14 &  10.0 & -0.08 & \ion{Si}{x} 50.6910 \\
\hline
\end{tabular}
\end{scriptsize}
\end{table*}
\setcounter{table}{4}
\begin{table*}
\caption{(cont). Chandra/LETG line fluxes of $\epsilon$~Eri$^a$}
\tabcolsep 3.pt
\begin{scriptsize}
\begin{tabular}{lrcrrrl}
\hline \hline
 Ion & {$\lambda$$_{\mathrm {model}}$} & 
 log $T_{\mathrm {max}}$ & $F_{\mathrm {obs}}$ & S/N & ratio & Blends \\ 
\hline
\ion{Si}{xi} &  52.2980 & 6.2 & 1.98e-14 &   7.1 & -0.12 & \ion{Si}{xi} 52.2690 \\
\ion{Fe}{xv} &  52.9110 & 6.4 & 2.24e-14 &   7.6 &  0.32 &  \\
\ion{Fe}{xv} &  53.1100 & 6.4 & 5.83e-15 &   3.9 &  0.21 &  \\
\ion{Fe}{xvi} &  54.1420 & 6.5 & 2.97e-14 &   8.8 &  0.29 &  \\
\ion{Fe}{xvi} &  54.7280 & 6.5 & 4.66e-14 &  11.0 &  0.19 & \ion{Fe}{xvi} 54.7630 \\
No id. &  56.9000 & \ldots & 2.61e-14 &  11.4 & \ldots & (\ion{O}{viii} 18.97, 3$^{\rm rd}$ order) \\
\ion{Mg}{x} &  57.8760 & 6.1 & 3.35e-14 &  12.8 &  0.09 & \ion{Fe}{xvii} 57.9119, \ion{Mg}{x} 57.9200 \\
No id. &  58.9500 & \ldots & 1.45e-14 &   8.3 & \ldots &  \\
\ion{Fe}{xv} &  59.4050 & 6.3 & 2.19e-14 &  10.2 & -0.07 &  \\
\ion{Fe}{xvi} &  62.8790 & 6.5 & 2.21e-14 &   7.2 & -0.20 &  \\
\ion{Mg}{x} &  63.1520 & 6.1 & 1.03e-14 &   4.9 & -0.10 &  \\
\ion{Mg}{x} &  63.2950 & 6.1 & 2.59e-14 &   7.8 &  0.00 & \ion{Mg}{x} 63.3110 \\
\ion{Fe}{xvi} &  63.7190 & 6.5 & 5.73e-14 &  11.6 & -0.10 &  \\
\ion{Fe}{xv} &  63.9580 & 6.3 & 1.13e-14 &   5.1 &  0.07 &  \\
\ion{Mg}{x} &  65.6730 & 6.1 & 9.06e-15 &   6.5 &  0.07 & \ion{Ne}{x} 65.6440 \\
\ion{Mg}{x} &  65.8450 & 6.1 & 1.26e-14 &   7.8 & -0.05 & \ion{Ne}{viii} 65.8940 \\
\ion{Fe}{xvi} &  66.2630 & 6.5 & 1.93e-14 &   9.4 & -0.10 &  \\
No id. &  66.3200 & \ldots & 1.45e-14 &   8.4 & \ldots & (\ion{O}{vii} 22.0977, 3$^{\rm rd}$ order) \\
\ion{Fe}{xvi} &  66.3680 & 6.5 & 3.60e-14 &  13.0 & -0.00 &  \\
\ion{Fe}{xv} &  69.6490 & 6.3 & 3.78e-14 &  13.5 & -0.25 &  \\
\ion{Fe}{xv} &  70.0540 & 6.3 & 1.94e-14 &   9.7 &  0.24 & \ion{Fe}{xv} 69.9450, 69.9870 \\
No id. &  71.9090 & \ldots & 1.27e-14 &   5.3 & \ldots & \ion{Mg}{ix} 71.901 \\
\ion{Mg}{ix} &  72.3120 & 6.0 & 1.10e-14 &   4.9 &  0.23 &  \\
\ion{Fe}{xv} &  73.4730 & 6.3 & 2.59e-14 &  11.0 &  0.00 & \ion{Ne}{viii} 73.4750 \\
No id. &  76.51700 & \ldots & 1.47e-14 &   8.1 & \ldots & \ion{Fe}{xvi} 76.51 \\
\ion{Ne}{viii} &  88.0820 & 5.8 & 4.25e-14 &  12.9 & -0.07 & \ion{Ne}{viii} 88.0820, 88.1190 \\
\ion{Fe}{xviii} &  93.9230 & 6.8 & 6.53e-14 &  16.0 & -0.23 &  \\
\ion{Ne}{viii} &  98.1150 & 5.8 & 1.44e-14 &   7.2 & -0.01 &  \\
\ion{Ne}{viii} &  98.2600 & 5.8 & 2.59e-14 &   9.6 & -0.05 & \ion{Ne}{viii} 98.2740 \\
\ion{Fe}{xix} & 101.5500 & 6.9 & 1.03e-14 &   6.0 & -0.33 &  \\
\ion{Fe}{xviii} & 103.9370 & 6.8 & 1.83e-14 &   8.0 & -0.30 &  \\
\ion{Fe}{xix} & 108.3700 & \ldots & 1.70e-14 &   7.7 & \ldots &  \\
\ion{Fe}{xxii} & 117.1700 & \ldots & 5.72e-15 &   4.5 & \ldots &  \\
\ion{Fe}{xx} & 121.8300 &\ldots  & 4.77e-15 &   3.9 & \ldots &  \\
\ion{Fe}{xx} & 132.8500 & \ldots & 2.94e-14 &   8.2 & \ldots &  \\
\ion{Ca}{xii} & 141.0380 & 6.3 & 1.61e-14 &   4.8 & -0.27 & \ion{Ca}{xii} 141.0380 \\
\ion{Fe}{ix} & 171.0730 & 5.9 & 1.45e-13 &  10.8 &  0.06 &  \\
\ion{Fe}{x} & 174.5340 & 6.0 & 7.62e-14 &   7.3 & -0.13 &  \\
\hline
\end{tabular}

{$^a$ Line fluxes (in erg cm$^{-2}$ s$^{-1}$) 
  measured in Chandra/LETG $\epsilon$~Eri spectrum. 
  log $T_{\mathrm {max}}$ indicates the maximum
  temperature (K) of formation of the line (unweighted by the
  EMD). ``Ratio'' is the log($F_{\mathrm {obs}}$/$F_{\mathrm {pred}}$) 
  of the line. 
  Blends amounting to more than 5\% of the total flux for each line are
  indicated. The lines in the range $\lambda$$\lambda$108--133 were excluded from the fit due to suspected problems of calibration.}
\end{scriptsize}
\end{table*}

\begin{table*}
\caption{Chandra/HETG line fluxes of $\lambda$~And$^a$}\label{tab:flland}
\tabcolsep 3.pt
\begin{scriptsize}
\begin{tabular}{lrcrrrl}
\hline \hline
 Ion & {$\lambda$$_{\mathrm {model}}$} &  
 log $T_{\mathrm {max}}$ & $F_{\mathrm {obs}}$ & S/N & ratio & Blends \\ 
\hline
\ion{Ar}{xvii} &   3.9491 & 7.3 & 2.09e-14 &   3.9 &  0.00 &  \\
\ion{S }{xvi} &   3.9908 & 7.4 & 9.47e-15 &   2.7 & -0.13 & \ion{S }{xvi}  3.9920, \ion{Ar}{xvii}  3.9941,  3.9942, \ion{S }{xv}  3.9980 \\
\ion{S }{xvi} &   4.7274 & 7.4 & 1.84e-14 &   3.7 & -0.09 & \ion{S }{xvi}  4.7328 \\
\ion{S }{xv} &   5.0387 & 7.2 & 4.05e-14 &   5.5 & -0.05 &  \\
\ion{S }{xv} &   5.0665 & 7.1 & 1.87e-14 &   3.7 &  0.23 & \ion{S }{xv}  5.0631 \\
\ion{S }{xv} &   5.1015 & 7.2 & 2.37e-14 &   4.2 &  0.06 & \ion{S }{xv}  5.0983, 5.1025 \\
\ion{Si}{xiv} &   5.2168 & 7.2 & 2.27e-14 &   3.9 &  0.01 & \ion{Si}{xiv}  5.2180 \\
\ion{Si}{xiv} &   6.1804 & 7.2 & 1.47e-13 &  20.4 &  0.01 & \ion{Si}{xiv}  6.1858 \\
\ion{Mg}{xii} &   6.5800 & 7.0 & 4.39e-15 &   4.2 & -0.23 & \ion{Fe}{xxiv}  6.5772, \ion{Mg}{xii}  6.5802 \\
\ion{Si}{xiii} &   6.6479 & 7.0 & 1.87e-13 &  25.7 & -0.02 &  \\
\ion{Si}{xiii} &   6.6882 & 7.0 & 3.16e-14 &  10.5 & -0.07 & \ion{Si}{xiii}  6.6850 \\
No id. &   6.7120 & \ldots & 2.42e-15 &   3.3 & \ldots &  \\
\ion{Si}{xiii} &   6.7403 & 7.0 & 1.23e-13 &  21.6 &  0.08 & \ion{Mg}{xii}  6.7378, \ion{Si}{xiii}  6.7432 \\
\ion{Mg}{xii} &   7.1058 & 7.0 & 4.53e-14 &  14.1 &  0.01 & \ion{Mg}{xii}  7.1069 \\
\ion{Al}{xiii} &   7.1710 & 7.1 & 3.22e-14 &  12.0 & -0.01 & \ion{Fe}{xxiv}  7.1690, \ion{Al}{xiii}  7.1764 \\
\ion{Mg}{xi} &   7.4730 & 6.8 & 1.07e-14 &   7.0 &  0.04 &  \\
\ion{Al}{xii} &   7.7573 & 6.9 & 3.04e-14 &  11.4 &  0.22 &  \\
\ion{Mg}{xi} &   7.8503 & 6.8 & 2.94e-14 &  11.4 &  0.01 &  \\
\ion{Al}{xii} &   7.8721 & 6.9 & 1.04e-14 &   6.7 & -0.12 & \ion{Al}{xii}  7.8721 \\
\ion{Fe}{xxiv} &   7.9857 & 7.3 & 1.56e-14 &   8.7 &  0.11 & \ion{Fe}{xxiv}  7.9960 \\
\ion{Fe}{xxiii} &   8.3038 & 7.2 & 8.09e-15 &   6.3 & -0.14 &  \\
\ion{Fe}{xxiv} &   8.3761 & 7.3 & 5.79e-15 &   5.3 &  0.17 & \ion{Fe}{xx}  8.3641 \\
\ion{Mg}{xii} &   8.4192 & 7.0 & 3.33e-13 &  40.2 &  0.06 & \ion{Mg}{xii}  8.4246 \\
\ion{Fe}{xxi} &   8.5740 & 7.1 & 8.48e-15 &   5.6 & -0.01 & \\
\ion{Fe}{xxii} &   8.9748 & 7.1 & 1.21e-14 &   6.9 & -0.06 &  \\
\ion{Ni}{xxvi} &   9.0603 & 7.4 & 1.33e-14 &   7.7 & -0.02 & \ion{Fe}{xxii}  9.0614, \ion{Fe}{xx}  9.0647, 9.0659,  9.0683 \\
No id. &   9.1300 & \ldots  & 1.15e-14 &   7.1 & \ldots &  \\
\ion{Mg}{xi} &   9.1687 & 6.8 & 2.03e-13 &  30.0 &  0.02 &  \\
\ion{Fe}{xxi} &   9.1944 & 7.1 & 1.69e-14 &   8.6 &  0.15 & \ion{Fe}{xx}  9.1882, 9.1979, \ion{Mg}{xi}  9.1927, 9.1938 \\
No id. &   9.2190 & \ldots  & 6.70e-15 &   5.4 & \ldots & \ion{Ne}{x} 9.215 \\
\ion{Mg}{xi} &   9.2312 & 6.8 & 3.27e-14 &  12.0 & -0.03 & \ion{Mg}{xi}  9.2282 \\
\ion{Ni}{xix} &   9.2540 & 6.9 & 7.45e-15 &   5.7 &  0.29 &  \\
No id. &   9.2870 & \ldots & 1.40e-14 &   7.8 & \ldots &  \ion{Ne}{x} 9.291 \\
\ion{Mg}{xi} &   9.3143 & 6.8 & 9.75e-14 &  20.6 &  0.09 &  \\
No id. &  9.3600 & \ldots  & 1.16e-14 &   7.1 & \ldots &  \\
\ion{Ni}{xx} &   9.3850 & 7.0 & 4.18e-15 &   4.2 & -0.29 & \ion{Ni}{xxvi}  9.3853, \ion{Ni}{xxv}  9.3900, \ion{Fe}{xxii}  9.3933 \\
\ion{Fe}{xxi} &   9.4797 & 7.0 & 3.71e-14 &  12.6 & -0.08 & \ion{Ne}{x}  9.4807, 9.4809 \\
\ion{Fe}{xix} &   9.6937 & 6.9 & 1.75e-14 &   8.5 &  0.42 & \ion{Fe}{xix}  9.6938 \\
\ion{Ne}{x} &   9.7080 & 6.8 & 5.33e-14 &  14.8 & -0.06 & \ion{Fe}{xix}  9.6938, \ion{Ne}{x}  9.7085 \\
\ion{Ni}{xix} &   9.9770 & 6.9 & 2.34e-14 &   9.6 & -0.37 & \ion{Ni}{xxv}  9.9700, \ion{Fe}{xxi}  9.9887, \ion{Fe}{xx}  9.9977, 10.0004, 10.0054 \\
No id. &  10.0200 & \ldots  & 1.89e-14 &   8.6 & \ldots & \ion{Na}{xi}
10.0232, 10.0286  \\
\ion{Fe}{xx} &  10.0529 & 7.0 & 1.16e-14 &   6.7 &  0.03 &  \\
\ion{Fe}{xx} &  10.1203 & 7.0 & 1.31e-14 &   7.1 & -0.18 & \ion{Ni}{xix} 10.1100, \ion{Fe}{xix} 10.1195, \ion{Fe}{xvii} 10.1210 \\
\ion{Ne}{x} &  10.2385 & 6.8 & 1.91e-13 &  26.9 &  0.12 & \ion{Ne}{x} 10.2396 \\
\ion{Fe}{xxiv} &  10.6190 & 7.3 & 3.43e-14 &  11.0 & -0.16 & \ion{Fe}{xix} 10.6295 \\
\ion{Fe}{xix} &  10.6491 & 6.9 & 1.34e-14 &   6.8 & -0.22 & \ion{Fe}{xix} 10.6414 \\
\ion{Fe}{xxiv} &  10.6630 & 7.3 & 2.28e-14 &   8.9 & -0.14 & \ion{Fe}{xvii} 10.6570 \\
\ion{Fe}{xvii} &  10.7700 & 6.8 & 1.32e-14 &   6.7 & -0.25 & \ion{Ne}{ix} 10.7650, \ion{Fe}{xix} 10.7650 \\
\ion{Fe}{xix} &  10.8160 & 6.9 & 2.17e-14 &   8.6 &  0.01 & \\
\ion{Fe}{xxiii} &  10.9810 & 7.2 & 3.23e-14 &  10.3 & -0.16 &  \\
\ion{Ne}{ix} &  11.0010 & 6.6 & 1.55e-14 &   7.1 & -0.03 & \ion{Ni}{xxii} 10.9920, 10.9927, \ion{Fe}{xix} 11.0022 \\
\ion{Fe}{xxiv} &  11.0290 & 7.3 & 4.17e-14 &  11.7 & -0.20 & \ion{Fe}{xxiii} 11.0190, \ion{Fe}{xvii} 11.0260 \\
No id. &  11.1000 & \ldots  & 6.16e-15 &   4.4 & \ldots &  \\
\ion{Fe}{xvii} &  11.1310 & 6.8 & 1.20e-14 &   6.2 & -0.28 & \ion{Fe}{xxii} 11.1376 \\
\ion{Fe}{xxiv} &  11.1760 & 7.3 & 3.96e-14 &  11.1 & -0.17 & \ion{Fe}{xxiv} 11.1870 \\
\ion{Fe}{xvii} &  11.2540 & 6.8 & 1.89e-14 &   7.6 & -0.35 & \ion{Fe}{xxiv} 11.2680 \\
\ion{Ni}{xxii} &  11.3049 & 7.1 & 9.47e-15 &   5.4 & -0.16 & \ion{Fe}{xviii} 11.2930, \ion{Fe}{xix} 11.2980, \ion{Ni}{xxi} 11.3180 \\
\ion{Fe}{xviii} &  11.3260 & 6.9 & 2.64e-14 &   8.9 & -0.13 & \ion{Fe}{xxiii} 11.3360 \\
\ion{Fe}{xviii} &  11.4230 & 6.9 & 2.48e-14 &   8.4 & -0.24 & \ion{Fe}{xxii} 11.4270 \\
\ion{Fe}{xxiv} &  11.4320 & 7.3 & 2.79e-14 &   8.9 & -0.00 &  \\
\ion{Fe}{xxiii} &  11.4580 & 7.2 & 1.08e-14 &   5.6 &  0.06 & \\
\ion{Fe}{xxii} &  11.4900 & 7.1 & 1.30e-14 &   6.1 & -0.19 & \\
\ion{Fe}{xviii} &  11.5270 & 6.9 & 2.62e-14 &   8.4 & -0.06 & \\
\ion{Ne}{ix} &  11.5440 & 6.6 & 3.90e-14 &  10.2 &  0.01 & \ion{Fe}{xviii} 11.5740 \\
\ion{Fe}{xxiii} &  11.7360 & 7.2 & 6.52e-14 &  12.9 & -0.16 &  \\
\ion{Fe}{xxii} &  11.7700 & 7.1 & 6.53e-14 &  13.2 & -0.19 &  \\
\ion{Fe}{xxii} &  11.8020 & 7.1 & 1.94e-14 &   7.2 &  0.28 & \\
\ion{Ni}{xx} &  11.8320 & 7.0 & 1.69e-14 &   6.7 &  0.08 &  \\
\ion{Fe}{xxii} &  11.9320 & 7.1 & 2.00e-14 &   7.3 &  0.07 &  \\
\ion{Fe}{xxii} &  11.9770 & 7.1 & 2.44e-14 &   8.0 & -0.11 & \ion{Fe}{xxi} 11.9750 \\
\ion{Ne}{x} &  12.1321 & 6.8 & 8.32e-13 &  45.9 & -0.06 & \ion{Fe}{xvii} 12.1240, \ion{Ne}{x} 12.1375 \\
\ion{Fe}{xxiii} &  12.1610 & 7.2 & 2.79e-14 &   8.4 & -0.26 &  \\
\ion{Fe}{xxii} &  12.2100 & 7.1 & 2.48e-14 &   7.9 &  0.18 &  \\
\ion{Fe}{xvii} &  12.2660 & 6.8 & 6.12e-14 &  12.3 & -0.03 &  \\
\ion{Fe}{xxi} &  12.2840 & 7.0 & 1.60e-13 &  19.8 &  0.01 &  \\
\ion{Fe}{xxi} &  12.3930 & 7.0 & 3.89e-14 &   9.7 &  0.11 & \ion{Fe}{xxi} 12.3940 \\
\ion{Ni}{xix} &  12.4350 & 6.9 & 4.83e-14 &  10.7 & -0.34 & \ion{Fe}{xxi} 12.4220, \ion{Fe}{xxii} 12.4311, 12.4318 \\
\ion{Fe}{xx} &  12.4680 & 7.0 & 1.52e-14 &   6.0 &  0.26 & \ion{Fe}{xxi} 12.4687 \\
\ion{Fe}{xxi} &  12.4990 & 7.0 & 4.18e-14 &   9.9 &  0.21 & \ion{Fe}{xxi} 12.4956 \\
\ion{Fe}{xxii} &  12.7540 & 7.1 & 3.33e-14 &   8.6 & -0.04 &  \\
No id. &  12.8100 & \ldots  & 6.88e-14 &  12.3 & \ldots &  \\
\ion{Fe}{xx} &  12.8240 & 7.0 & 1.66e-13 &  19.0 & -0.29 & \ion{Fe}{xxi} 12.8220, \ion{Fe}{xx} 12.8460, 12.8640 \\
\hline
\end{tabular}
\end{scriptsize}
\end{table*}
\setcounter{table}{5}
\begin{table*}
\caption{(cont). Chandra/HETG line fluxes of $\lambda$~And$^a$}
\tabcolsep 3.pt
\begin{scriptsize}
\begin{tabular}{lrcrrrl}
\hline \hline
 Ion & {$\lambda$$_{\mathrm {model}}$} & 
 log $T_{\mathrm {max}}$ & $F_{\mathrm {obs}}$ & S/N & ratio & Blends \\ 
\hline
\ion{Fe}{xx} &  12.9650 & 7.0 & 9.34e-14 &  14.1 & -0.10 & \ion{Fe}{xx} 12.9120, \ion{Fe}{xix} 12.9311, 12.9330, \ion{Fe}{xxii} 12.9530 \\
\ion{Fe}{xx} &  12.9920 & 7.0 & 1.56e-14 &   5.7 & -0.19 & \\
\ion{Fe}{xix} &  13.0220 & 6.9 & 3.26e-14 &   8.2 &  0.05 & \ion{Fe}{xx} 13.0240 \\
\ion{Fe}{xx} &  13.0610 & 7.0 & 3.61e-14 &   8.7 &  0.01 & \ion{Fe}{xx} 13.0580 \\
\ion{Fe}{xx} &  13.1000 & 7.0 & 1.35e-14 &   5.2 & -0.05 &  \\
\ion{Fe}{xx} &  13.1530 & 7.0 & 4.54e-14 &   9.6 & -0.09 & \ion{Fe}{xx} 13.1370 \\
\ion{Fe}{xx} &  13.2740 & 7.0 & 5.72e-14 &  10.6 &  0.05 & \ion{Fe}{xx} 13.2240, 13.2241, \ion{Fe}{xxii} 13.2360, \ion{Fe}{xxi} 13.2487, \ion{Ni}{xx} 13.2560, \ion{Fe}{xix} 13.2658 \\
\ion{Ni}{xx} &  13.3090 & 6.9 & 3.99e-14 &   8.8 &  0.04 & \ion{Fe}{xx} 13.3017, 13.3089, \ion{Fe}{xix} 13.3191, \ion{Fe}{xviii} 13.3230, \ion{Ni}{xx} 13.3236 \\
\ion{Fe}{xx} &  13.3850 & 7.0 & 4.82e-14 &   9.6 &  0.03 & \ion{Fe}{xviii} 13.3550, 13.3948 \\
\ion{Fe}{xx} &  13.4090 & 7.0 & 2.51e-14 &   6.9 &  0.27 & \ion{Fe}{xviii} 13.4070 \\
\ion{Fe}{xix} &  13.4230 & 6.9 & 4.01e-14 &   8.7 &  0.19 & \ion{Fe}{xx} 13.4181, \ion{Fe}{xxi} 13.4308 \\
\ion{Ne}{ix} &  13.4473 & 6.6 & 2.84e-13 &  23.2 &  0.02 & \ion{Fe}{xix} 13.4620 \\
\ion{Fe}{xxi} &  13.5070 & 7.0 & 9.92e-14 &  13.6 &  0.04 & \ion{Fe}{xix} 13.4970 \\
\ion{Fe}{xix} &  13.5180 & 6.9 & 9.04e-14 &  13.0 & -0.14 &  \\
\ion{Fe}{xx} &  13.5350 & 7.0 & 3.75e-14 &   8.3 &  0.28 & \\
\ion{Ne}{ix} &  13.5531 & 6.6 & 3.67e-14 &   8.2 & -0.13 & \ion{Fe}{xix} 13.5510, 13.5540, \ion{Fe}{xx} 13.5583 \\
\ion{Fe}{xix} &  13.6450 & 6.9 & 3.83e-14 &   8.1 &  0.18 & \ion{Fe}{xix} 13.6481 \\
\ion{Fe}{xix} &  13.6828 & 6.9 & 5.86e-14 &  10.1 &  0.35 & \ion{Fe}{xix} 13.6742, 13.6752, 13.6881, 13.6897 \\
\ion{Ne}{ix} &  13.6990 & 6.6 & 1.10e-13 &  13.8 &  0.06 &  \\
\ion{Fe}{xix} &  13.7950 & 6.9 & 1.35e-13 &  15.2 & -0.08 & \ion{Fe}{xix} 13.7315, 13.7458, 13.7590, \ion{Fe}{xx} 13.7670, \ion{Ni}{xix} 13.7790 \\
\ion{Fe}{xvii} &  13.8250 & 6.8 & 4.80e-14 &   9.0 &  0.02 &  \\
\ion{Fe}{xix} &  13.8390 & 6.9 & 1.72e-14 &   5.4 & -0.14 & \ion{Fe}{xx} 13.8430 \\
\ion{Fe}{xviii} &  13.9530 & 6.9 & 2.93e-14 &   6.7 & -0.08 & \ion{Fe}{xix} 13.9549, 13.9551, \ion{Fe}{xx} 13.9620 \\
\ion{Fe}{xxi} &  14.0080 & 7.0 & 2.52e-14 &   6.5 &  0.00 &  \\
\ion{Ni}{xix} &  14.0430 & 6.8 & 5.85e-14 &   9.8 &  0.24 &  \\
\ion{Ni}{xix} &  14.0770 & 6.8 & 3.17e-14 &   7.2 &  0.02 &  \\
\ion{Fe}{xviii} &  14.2080 & 6.9 & 1.80e-13 &  16.9 & -0.18 & \\
\ion{Fe}{xviii} &  14.2560 & 6.9 & 8.16e-14 &  10.7 & -0.04 & \ion{Fe}{xx} 14.2670 \\
\ion{Fe}{xviii} &  14.3430 & 6.9 & 1.89e-14 &   4.8 & -0.29 & \\
\ion{Fe}{xviii} &  14.3730 & 6.9 & 6.08e-14 &   8.5 & -0.05 &  \\
\ion{Fe}{xviii} &  14.5340 & 6.9 & 5.80e-14 &   8.4 &  0.06 &  \\
\ion{Fe}{xviii} &  14.5710 & 6.9 & 7.69e-15 &   3.1 & -0.54 & \ion{Fe}{xx} 14.5617 \\
\ion{Fe}{xviii} &  14.6011 & 6.8 & 1.23e-14 &   3.8 & -0.04 & \ion{Fe}{xviii} 14.6006 \\
\ion{Fe}{xix} &  14.6640 & 6.9 & 4.63e-14 &   7.6 &  0.11 &  \\
\ion{Fe}{xix} &  14.7250 & 6.9 & 1.93e-14 &   5.1 & -0.05 & \ion{Fe}{xviii} 14.7260 \\
\ion{Fe}{xx} &  14.7540 & 7.0 & 2.21e-14 &   5.7 &  0.02 &  \\
\ion{O }{viii} &  14.8205 & 6.5 & 3.51e-14 &   7.6 &  0.20 &  \\
\ion{Fe}{xx} &  14.9703 & 7.0 & 3.77e-14 &   8.1 &  0.11 & \ion{Fe}{xix} 14.9610 \\
\ion{Fe}{xvii} &  15.0140 & 6.7 & 2.95e-13 &  22.5 & -0.26 &  \\
\ion{Fe}{xx} &  15.0470 & 7.0 & 2.25e-14 &   6.2 &  0.27 & \\
\ion{Fe}{xix} &  15.0790 & 6.9 & 6.87e-14 &  10.8 &  0.23 &  \\
\ion{O }{viii} &  15.1760 & 6.5 & 6.76e-14 &  10.6 & -0.08 & \ion{O }{viii} 15.1765, \ion{Fe}{xix} 15.1770 \\
\ion{Fe}{xix} &  15.1980 & 6.9 & 5.37e-14 &   9.5 &  0.20 &  \\
\ion{Fe}{xvii} &  15.2610 & 6.7 & 1.33e-13 &  14.8 & -0.07 &  \\
\ion{Fe}{xvii} &  15.4530 & 6.7 & 2.93e-14 &   6.8 &  0.23 &  \\
\ion{Fe}{xviii} &  15.6250 & 6.8 & 7.00e-14 &  10.4 &  0.03 &  \\
\ion{Fe}{xviii} &  15.8240 & 6.8 & 5.04e-14 &   8.7 &  0.11 &  \\
\ion{Fe}{xviii} &  15.8700 & 6.8 & 4.68e-14 &   8.3 &  0.32 &  \\
\ion{O }{viii} &  16.0055 & 6.5 & 2.37e-13 &  18.6 & -0.11 & \ion{Fe}{xviii} 16.0040, \ion{O }{viii} 16.0067 \\
\ion{Fe}{xviii} &  16.0710 & 6.8 & 1.54e-13 &  14.9 &  0.25 &  \\
\ion{Fe}{xix} &  16.1100 & 6.9 & 4.77e-14 &   8.3 & -0.01 & \ion{Fe}{xviii} 16.1127 \\
\ion{Fe}{xviii} &  16.1590 & 6.8 & 1.52e-14 &   4.7 & -0.37 &  \\
\ion{Fe}{xvii} &  16.2285 & 6.7 & 8.52e-15 &   3.5 &  0.24 &  \\
\ion{Fe}{xix} &  16.2830 & 6.9 & 8.24e-15 &   3.4 & -0.28 & \ion{Fe}{xviii} 16.2882 \\
\ion{Fe}{xviii} &  16.3200 & 6.8 & 1.75e-14 &   5.0 &  0.26 &  \\
\ion{Fe}{xvii} &  16.3500 & 6.7 & 1.57e-14 &   4.7 &  0.18 & \ion{Fe}{xix} 16.3414 \\
\ion{Fe}{xvii} &  16.7800 & 6.7 & 1.74e-13 &  15.0 & -0.12 &  \\
\ion{Fe}{xvii} &  17.0510 & 6.7 & 2.41e-13 &  17.4 & -0.01 &  \\
\ion{Fe}{xvii} &  17.0960 & 6.7 & 2.25e-13 &  16.7 &  0.06 &  \\
\ion{Fe}{xviii} &  17.6230 & 6.8 & 4.68e-14 &   7.1 & -0.08 &  \\
\ion{O }{vii} &  18.6270 & 6.3 & 1.01e-14 &   3.1 & -0.31 & \ion{Ar}{xvi} 18.6240 \\
\ion{Ca}{xviii} &  18.6910 & 6.9 & 7.87e-15 &   2.7 & -0.00 &  \\
\ion{O }{viii} &  18.9671 & 6.5 & 1.00e-12 &  29.4 & -0.09 & \ion{O }{viii} 18.9725 \\
\ion{O }{vii} &  21.6015 & 6.3 & 1.30e-13 &   8.2 &  0.03 &  \\
\ion{O }{vii} &  21.8036 & 6.3 & 1.01e-14 &   2.2 & -0.22 &  \\
\ion{O }{vii} &  22.0977 & 6.3 & 7.63e-14 &   5.7 &  0.05 & \ion{Ca}{xvii} 22.1140 \\
\hline
\end{tabular}

{$^a$ Line fluxes (in erg cm$^{-2}$ s$^{-1}$) 
  measured in Chandra/HETG $\lambda$~And spectra. 
  log $T_{\mathrm {max}}$ indicates the maximum
  temperature (K) of formation of the line (unweighted by the
  EMD). ``Ratio'' is the log($F_{\mathrm {obs}}$/$F_{\mathrm {pred}}$) 
  of the line. 
  Blends amounting to more than 5\% of the total flux for each line are
  indicated.}
\end{scriptsize}
\end{table*}

\begin{table*}
\caption{XMM/RGS line fluxes of V851 Cen$^a$}\label{tab:flv851}
\tabcolsep 3.pt
\begin{scriptsize}
\begin{tabular}{lrcrrrl}
\hline \hline
 Ion & {$\lambda$$_{\mathrm {model}}$} &  
 log $T_{\mathrm {max}}$ & $F_{\mathrm {obs}}$ & S/N & ratio & Blends \\ 
\hline
\ion{Si}{xiii} &  6.6480 & 7.0 & 4.51e-14 &   3.1 & -0.00 & \ion{Si}{xiii}  6.6882, 6.7403, \ion{Mg}{xii}  6.7378 \\
\ion{Mg}{xii} &  8.4192 & 7.0 & 9.90e-14 &   7.8 &  0.07 & \ion{Mg}{xii}  8.4246 \\
\ion{Mg}{xi} &  9.1687 & 6.8 & 3.52e-14 &   4.8 & -0.14 &  \\
\ion{Mg}{xi} &  9.2312 & 6.8 & 4.64e-14 &   4.1 &  0.10 & \ion{Mg}{xi}  9.3143, \ion{Ni}{xxv}  9.3400 \\
\ion{Fe}{xxi} &  9.4797 & 7.0 & 3.82e-14 &   4.2 &  0.27 & \ion{Ne}{x}  9.4807,  9.4809 \\
\ion{Ne}{x} &  9.7080 & 6.8 & 2.29e-14 &   4.1 & -0.20 & \ion{Ne}{x}  9.7085 \\
No id. & 10.0200 & \ldots & 4.41e-14 &   3.0 & \ldots & \ion{Na}{xi}
10.0232, 10.0286 \\
\ion{Ne}{x} & 10.2385 & 6.8 & 8.14e-14 &   7.2 & -0.06 & \ion{Ne}{x} 10.2396 \\
\ion{Fe}{xxiv} & 10.6190 & 7.3 & 3.80e-14 &   3.2 &  0.02 & \ion{Fe}{xix} 10.6414, 10.6491, \ion{Fe}{xvii} 10.6570, \ion{Fe}{xxiv} 10.6630 \\
\ion{Fe}{xvii} & 10.7700 & 6.8 & 3.44e-14 &   2.4 &  0.09 & \ion{Ni}{xxiii} 10.7214, 10.8491, \ion{Ne}{ix} 10.7650, \ion{Fe}{xix} 10.8160 \\
\ion{Fe}{xxiii} & 11.0190 & 7.2 & 4.97e-14 &   3.2 & -0.07 & \ion{Fe}{xxiii} 10.9810, \ion{Ne}{ix} 11.0010, \ion{Fe}{xxiv} 11.0290 \\
\ion{Fe}{xxiv} & 11.1760 & 7.3 & 2.64e-14 &   4.8 & -0.04 & \ion{Fe}{xvii} 11.1310, \ion{Fe}{xxiv} 11.1870 \\
\ion{Fe}{xvii} & 11.2540 & 6.8 & 3.45e-14 &   2.9 &  0.25 & \ion{Ni}{xxii} 11.2118, \ion{Fe}{xxiv} 11.2680, \ion{Fe}{xxiii} 11.2850, \ion{Ni}{xxi} 11.2908 \\
\ion{Fe}{xviii} & 11.4230 & 6.9 & 2.20e-14 &   2.2 & -0.14 & \ion{Fe}{xxii} 11.4270, \ion{Fe}{xxiv} 11.4320, \ion{Fe}{xxiii} 11.4580 \\
\ion{Fe}{xviii} & 11.5270 & 6.9 & 4.40e-14 &   3.8 &  0.03 & \ion{Fe}{xxii} 11.4900, \ion{Ni}{xix} 11.5390, \ion{Ne}{ix} 11.5440 \\
\ion{Fe}{xxii} & 11.7700 & 7.1 & 8.27e-14 &   6.0 & -0.01 & \ion{Fe}{xxiii} 11.7360 \\
\ion{Ne}{x} & 12.1320 & 6.8 & 4.99e-13 &   5.9 &  0.12 & \ion{Ne}{x} 12.1321 \\
\ion{Fe}{xxi} & 12.2840 & 7.0 & 3.64e-14 &   3.2 & -0.18 & \ion{Fe}{xvii} 12.2660 \\
\ion{Ni}{xix} & 12.4350 & 6.9 & 3.97e-14 &   3.5 & -0.17 & \ion{Fe}{xxi} 12.3930, 12.4220, 12.4990 \\
\ion{Fe}{xx} & 12.8240 & 7.0 & 5.31e-14 &   5.2 & -0.13 & \ion{Fe}{xxi} 12.8220, \ion{Fe}{xx} 12.8460, 12.8640 \\
\ion{Fe}{xx} & 12.9650 & 7.0 & 4.00e-14 &   4.3 &  0.04 & \ion{Fe}{xx} 12.9120, 12.9920, 13.0240,\ion{Fe}{xix} 12.9330, 13.0220, \ion{Fe}{xxii} 12.9530  \\
\ion{Ne}{ix} & 13.4473 & 6.6 & 2.01e-13 &  13.3 &  0.12 &  \\
\ion{Fe}{xix} & 13.5180 & 6.9 & 5.92e-14 &   6.9 & -0.13 & \ion{Fe}{xix} 13.4970, \ion{Fe}{xxi} 13.5070, \ion{Ne}{ix} 13.5531 \\
\ion{Ne}{ix} & 13.6990 & 6.6 & 8.12e-14 &   8.5 &  0.05 & \ion{Fe}{xix} 13.6450, 13.7315, 13.7458 \\
\ion{Fe}{xix} & 13.7950 & 6.9 & 3.81e-14 &   4.2 & -0.04 & \ion{Fe}{xx} 13.7670, \ion{Ni}{xix} 13.7790, \ion{Fe}{xvii} 13.8250 \\
\ion{Fe}{xxi} & 14.0080 & 7.0 & 4.81e-14 &   5.7 &  0.07 & \ion{Ni}{xix} 14.0430, 14.0770 \\
\ion{Fe}{xviii} & 14.2080 & 6.9 & 6.40e-14 &   6.6 & -0.13 & \ion{Fe}{xviii} 14.2080, \ion{Fe}{xviii} 14.2560, \ion{Fe}{xx} 14.2670 \\
\ion{Fe}{xviii} & 14.3730 & 6.9 & 3.16e-14 &   5.0 & -0.15 & \ion{Fe}{xx} 14.3318, \ion{Fe}{xviii} 14.3430, \ion{Fe}{xviii} 14.3430, \ion{Fe}{xx} 14.4207, \ion{Fe}{xviii} 14.4250, \ion{Fe}{xviii} 14.4392 \\
\ion{Fe}{xviii} & 14.5340 & 6.9 & 4.32e-14 &   4.3 &  0.24 & \ion{Fe}{xviii} 14.4856, \ion{Fe}{xviii} 14.5056, \ion{Fe}{xviii} 14.5710, \ion{Fe}{xviii} 14.6011 \\
\ion{O }{viii} & 14.8205 & 6.5 & 3.01e-14 &   4.6 &  0.31 & \ion{Fe}{xviii} 14.7820, \ion{O }{viii} 14.8207, \ion{Fe}{xx} 14.8276 \\
\ion{Fe}{xvii} & 15.0140 & 6.7 & 9.47e-14 &  10.6 & -0.21 & \ion{Fe}{xix} 15.0790 \\
\ion{O }{viii} & 15.1760 & 6.5 & 5.19e-14 &   9.0 &  0.18 & \ion{O }{viii} 15.1765, \ion{Fe}{xix} 15.1980 \\
\ion{Fe}{xvii} & 15.2610 & 6.7 & 2.78e-14 &   4.4 & -0.11 &  \\
\ion{Fe}{xvii} & 15.4530 & 6.7 & 1.56e-14 &   3.8 &  0.25 & \ion{Fe}{xix} 15.4136, \ion{Fe}{xviii} 15.4940, \ion{Fe}{xx} 15.5170, \ion{Fe}{xviii} 15.5199 \\
\ion{Fe}{xviii} & 15.6250 & 6.8 & 2.62e-14 &   4.6 &  0.23 &  \\
\ion{Fe}{xviii} & 15.8700 & 6.8 & 2.02e-14 &   3.7 &  0.13 & \ion{Fe}{xviii} 15.8240 \\
\ion{O }{viii} & 16.0066 & 6.5 & 1.16e-13 &  10.3 & -0.07 & \ion{Fe}{xviii} 16.0040, \ion{O }{viii} 16.0055, \ion{Fe}{xviii} 16.0710, \ion{Fe}{xix} 16.1100, \ion{Fe}{xviii} 16.1590 \\
\ion{Fe}{xvii} & 16.7800 & 6.7 & 3.85e-14 &   6.1 & -0.15 &  \\
\ion{Fe}{xvii} & 17.0510 & 6.7 & 1.02e-13 &   7.3 & -0.01 & \ion{Fe}{xvii} 17.0960 \\
\ion{Fe}{xviii} & 17.6230 & 6.8 & 6.16e-15 &   1.7 & -0.34 &  \\
\ion{O }{vii} & 18.6270 & 6.3 & 1.83e-14 &   4.0 &  0.03 & \ion{Ca}{xviii} 18.6910 \\
\ion{O }{viii} & 18.9671 & 6.5 & 3.45e-13 &  27.5 & -0.08 & \ion{O }{viii} 18.9725 \\
\ion{Ca}{xvi} & 21.4500 & 6.7 & 3.57e-15 &   2.2 & -0.02 & \ion{Ca}{xvi} 21.4410 \\
\ion{O }{vii} & 21.6015 & 6.3 & 3.78e-14 &   4.2 & -0.08 & \ion{Ca}{xvi} 21.6100 \\
\ion{O }{vii} & 22.0977 & 6.3 & 1.39e-14 &   2.2 & -0.26 & \ion{Ca}{xvii} 22.1140 \\
\ion{Ar}{xvi} & 23.5460 & 6.7 & 6.92e-15 &   1.4 &  0.09 & \ion{Ar}{xvi} 23.5900, \ion{Ca}{xvi} 23.6260 \\
\ion{N }{vii} & 24.7792 & 6.3 & 3.26e-14 &   7.4 &  0.00 & \ion{N }{vii} 24.7846, \ion{Ar}{xvi} 24.8540 \\
\ion{Ar}{xvi} & 24.9910 & 6.7 & 4.30e-15 &   1.4 & -0.08 & \ion{Ar}{xvi} 24.9910, \ion{Ar}{xvi} 25.0130, \ion{Ar}{xv} 25.0500 \\
\ion{C }{vi} & 26.9896 & 6.2 & 2.70e-15 &   1.2 &  0.30 & \ion{C }{vi} 26.9901 \\
\ion{C }{vi} & 28.4652 & 6.2 & 2.35e-15 &   1.8 & -0.21 & \ion{C }{vi} 28.4663 \\
\ion{C }{vi} & 33.7342 & 6.1 & 2.10e-14 &   4.6 & -0.01 & \ion{C }{vi} 33.7396 \\
\hline
\end{tabular}

{$^a$ Line fluxes (in erg cm$^{-2}$ s$^{-1}$) 
  measured in XMM/RGS V851 Cen summed spectra. 
  log $T_{\rm max}$ indicates the maximum
  temperature (K) of formation of the line (unweighted by the
  EMD). ``Ratio'' is the log($F_{\mathrm {obs}}$/$F_{\mathrm {pred}}$) 
  of the line. 
  Blends amounting to more than 5\% of the total flux for each line are
  indicated.}
\end{scriptsize}
\end{table*}

\end{document}